\documentclass[aps,pra,twocolumn,showpacs]{revtex4}
 
\usepackage [latin1] {inputenc}
\usepackage {t1enc}
\usepackage [dvips] {graphicx}
\usepackage {verbatim}
\usepackage {epsfig}
\usepackage {here}
\usepackage {amsmath}
\usepackage {amssymb}
\usepackage {bbm}

\newcommand{\eqnref}[1]{(\ref{#1})}
\newcommand{\figref}[1]{Fig.~\ref{#1}}

\newcommand{\secref}[1]{Sec.~\ref{#1}}
\newcommand{\appref}[1]{App.~\ref{#1}}

\newcommand{\ah}{\hat a}

\newcommand{\vacr}{|vac\rangle}
\newcommand{\vacl}{\langle vac|}

\newcommand{\gF}{\mathcal{F}}
\newcommand{\bra}[1]{\langle #1|}
\newcommand{\ket}[1]{|#1\rangle}
\newcommand{\ndc}{\bar{n}_{dc}}
 
\makeindex

\begin{document}

\title{Memory Imperfections in Atomic Ensemble-based Quantum Repeaters}

\author{Jonatan Bohr Brask}\author{Anders Søndberg Sørensen}
\affiliation{QUANTOP, The Niels Bohr Institute, University of Copenhagen, 2100 Copenhagen Ø, Denmark}

\begin{abstract}
Quantum repeaters promise to deliver long-distance entanglement overcoming noise and loss in realistic quantum channels. A promising class of repeaters, based on atomic ensemble quantum memories and linear optics, follow the proposal by Duan \textit{et al} [Nature 414, 413, 2001]. Here we analyse this protocol in terms of a very general model for the quantum memories employed. We derive analytical expressions for scaling of entanglement with memory imperfections, dark counts, loss and distance, and apply our results to two specific quantum memory protocols. Our methods apply to any quantum memory with an interaction Hamiltonian at most quadratic in the mode operators and are in principle extendible to more recent modifications of the original DLCZ proposal.
\end{abstract}


\maketitle

\section{Introduction}

If one attempts to transfer quantum information by direct transmission, the communication rate decreases exponentially with distance, due to decoherence and loss. Quantum repeaters achieve subexponential scaling by generating entanglement locally in parallel for many segments of a short length $L_0$ and subsequently extending the distance by entanglement swapping until the full channel length $L=2^nL_0$ is reached \cite{briegel}. Quantum memories play a crucial role in quantum repeaters because operations take place on many segments in parallel and the operations in each segment may fail with a large probability. It is therefore essential to have a quantum memory where entanglement successfully generated in one segment may be stored, while entanglement generation and connection is being attempted in other segments. One promising repeater protocol based on storage of light in atomic ensembles was proposed by Duan, Lukin, Cirac and Zoller (DLCZ) \cite{dlcz} and later improved upon in a number of papers \cite{chen,jiang,simon,sangouard,sangouard2}. Extensive experimental progress have been made toward the realisation of this protocol \cite{chaneliere,chou,eisaman,chen2}.

In this paper, we consider the effect of memory imperfections in a repeater architecture closely resembling the original DLCZ proposal. Using a general model for memories, we investigate how the repeater performance depends on the memory properties in the presence of realistic errors, i.e. lossy fibres, detection inefficiency and dark counts. We then apply the results to specific ensemble based memories \cite{dlcz,muschik,julsgaard} and evaluate the performance of quantum repeaters based on these memories. The methods we develop, though applied here to the DLCZ architecture, could be extended also to the more recent protocols \cite{chen,jiang,simon,sangouard,sangouard2,collins}.


\section{Modelling the repeater}

\begin{figure}
\centering
\includegraphics[width=.4\textwidth]{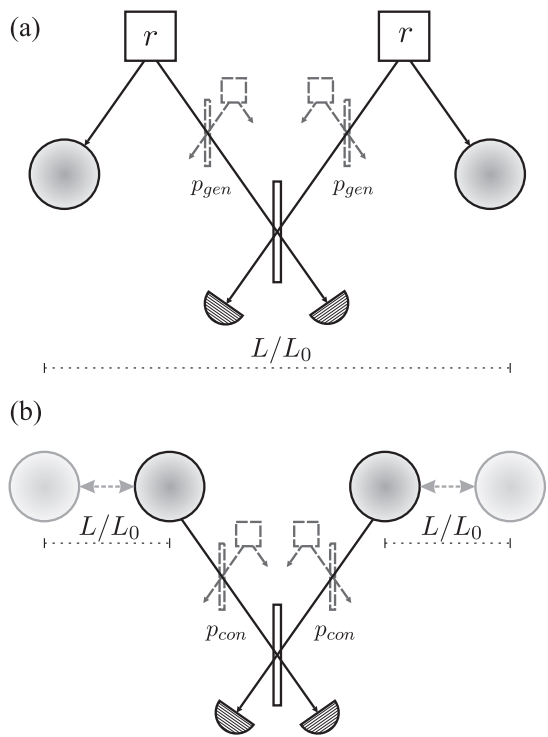}
\caption{\textbf{(a)} Entanglement generation. Two PDC's (large boxes) emit two-mode squeezed states. One mode from each pair is stored in an atomic memory (circles), while the remaining modes are mixed on a 50/50 beam splitter and measured. Observation of a single click leads to entanglement between the atomic modes. The dashed elements are virtual beam splitters and PDC's modelling losses and detector dark counts. \textbf{(b)} Entanglement connection is achieved by reading out the connecting ends of two entangled pairs and mixing on a 50/50 beam splitter. A single click heralds successful connection. For perfect memories, this setup is equivalent to the DLCZ setup \cite{dlcz}.}\label{fig.entgenandcon}
\end{figure}

The repeater that we shall consider is defined by the setups for entanglement generation and connection, illustrated in \figref{fig.entgenandcon}. Although the use of repeaters is motivated by the presence of errors in transmission, it is easier to understand the basics of the protocol in the absence of errors. Hence we consider this ideal case first.

To generate entanglement in one segment of the repeater, two non-degenerate parametric down converters (PDC's) are used. The two-mode squeezed state generated by a single PDC with small squeezing parameter $r$ is
\begin{equation}
\label{eq.twomodesqueeze}
\frac{1}{\cosh{r}} \sum_{k=0}^{\infty} (\tanh{r})^k \ket{k,k} \approx \vacr + r\ket{11} + O(r^2) ,
\end{equation}
where $\ket{k_1,k_2}$ denotes a Fock state with photons numbers $k_1$ and $k_2$ in modes 1 and 2. A single excitation is subtracted non-locally from the two squeezed pairs, by mixing one mode from each pair on a balanced beam splitter and conditioning on a single click (\figref{fig.entgenandcon}a). In the ideal case of noiseless operations and photon number resolving detectors, this leaves a Bell state to be stored in the atomic memories
\begin{equation}
\label{eq.psiplus}
\ket{\Psi^+} = \frac{1}{\sqrt{2}}(\ket{01} + \ket{10}) .
\end{equation}
To extend the entanglement distance, two neighbouring segments are connected by mixing one mode from each on a balanced beam splitter and conditioning on a single click (\figref{fig.entgenandcon}b). After the connection, the remaining modes are again in a Bell state, and thus under ideal conditions the protocol generates a maximally entangled pair over the distance $L$.

In Ref. \cite{dlcz} it was shown that the type of protocol above works also in the presence of photon loss. By conditioning on a click one essentially purifies the state from losses. Other errors, however, are not purified in this way. Our goal here is to make a detailed investigation of the effect of various types of noise on the efficiency of the protocol. The noise sources that we will consider are: (i) \textit{transmission losses}, (ii) \textit{detector dark counts} and (iii) \textit{memory imperfections}.

To model the transmission losses (i), virtual beam splitters are inserted into the setup as illustrated in \figref{fig.entgenandcon}. For simplicity we assume that the memory is located close to the PDC's, such that the loss in the memory arm is small and can be treated perturbatively as a memory imperfection. In the detector arms of the entanglement generation and connection setups, the loss probabilities are $p_{gen}$ and $p_{con}$ respectively. Photon loss due to detector inefficiency can be included in the transmission loss.

To model the dark counts (ii), we assume that the signal to be measured is mixed with a thermal state, which we generate by inserting a virtual PDC (see \figref{fig.entgenandcon}). We choose a thermal distribution for the dark-counts, because this is easily treated as a Bogoliubov transformation (see below). The repeater setup presented here is only feasible for $\bar{n}_{dc} \ll 1$, where $\bar{n}_{dc}$ is the average number of dark counts per detector in one measurement cycle. Therefore only the first-order contribution from dark counts is considered, and the actual distribution is not important. The reflection coefficients of the virtual beam splitters are given by the photon loss probabilities, and the detector dark count rate determines the squeezing parameter in the virtual PDC's (see \appref{app.moderedux}).
 
Memory imperfections (iii) are described by the physical parameters of the particular quantum memory inserted in the setup. However, it is desirable to analyse the problem using a general description of the memories. The incoming and outgoing light fields are conveniently described by harmonic oscillator degrees of freedom, and ideally the memories map the state of the incoming light to the outgoing light. In the Heisenberg picture this implies a mapping $\hat a '= \hat a$, where $\hat a '$ and $ \hat a$ are the field operators for the outgoing and incoming modes respectively. In the presence of imperfections, the state transfer will be described by an admixture of other field operators into the outgoing mode. Here we are mainly interested in describing quantum memories based on atomic ensembles. Such ensembles can in the limit of many atoms be described by a set of harmonic oscillators. Furthermore,  ensemble based quantum memories are to leading order typically described by interaction Hamiltonians which are quadratic in the field operators for the light and atoms \cite{dlcz, muschik, julsgaard}. The resulting state evolution can then be described by a Bogoliubov (i.e. linear, unitary) transformation of the field operators even in the presence of imperfections such as spontaneous emission. We shall consider quantum memories described by the most general possible Bogoliubov transformation and our results thus apply to a very wide class of quantum
memories. Our calculations cannot, however, describe interactions with Hamiltonians of higher order than quadratic in the mode operators. E.g. we cannot describe the optical Kerr effect or or single-atom memories.

In addition to the errors (i),(ii),(iii) we also consider detectors which do not resolve the photon number, as efficient, single-photon counters are difficult to construct and it is interesting to compare the two cases of counting/non-counting. We note that the DLCZ setup is equivalent to the protocol considered here in the case of perfect memories, i.e. in the absence of type (iii) noise \cite{dlcz}. The results presented below therefore also apply to that protocol. 

\begin{figure}
\centering
\includegraphics[width=.35\textwidth]{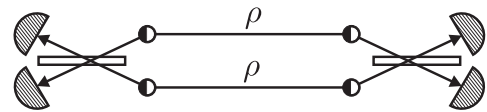}
\caption{Given two copies of a single-rail entangled state $\rho$, dual-rail entanglement is obtained by mixing and conditioning on a single click at each end. The half-filled circles denote excitations shared between two modes.}\label{fig.bellparamdef}
\end{figure}

The figures of merit for a repeater are the \textit{rate} $R$ at which entangled pairs are generated and the quality of the entanglement as functions of the distance $L$. As seen below, and as noted in Ref. \cite{dlcz}, for lossy or non-resolving detectors the final state will contain a large vacuum component with no excitations in the ensembles. However, by combining two single-rail qubits into one dual-rail qubit via postselection as shown in \figref{fig.bellparamdef} the vacuum component can be removed and interesting applications such as quantum cryptography or teleportation can be performed provided that the quality of entanglement conditioned on successful postselection is high \cite{dlcz,simon}.

As a measure of entanglement, we use the Bell parameter $S$. Let $P_{diff}(\phi,\varphi)$ ($P_{same}(\phi,\varphi)$) denote the conditional probability that one upper and one lower detector (both upper or both lower detectors) on \figref{fig.bellparamdef} click, given that a single click is detected at each end. We define $E(\phi,\varphi) = P_{same}(\phi,\varphi) - P_{diff}(\phi,\varphi)$ and
\begin{equation}
S = E(\frac{\pi}{2},\frac{\pi}{4}) + E(0,\frac{\pi}{4}) + E(0,\frac{-\pi}{4}) - E(\frac{\pi}{2},\frac{-\pi}{4}) .
\label{eq.sdef}
\end{equation}
The parameter space is then $-2\sqrt{2} \leq S \leq 2\sqrt{2}$ with $|S| > 2$ designating that the state is entangled \cite{aspect}. Note that for ease of computation we have fixed the angles in the definition of $S$, whereas the custom definition is a maximum over varying angles. The angles in \eqnref{eq.sdef} maximize $S$ for the ideal state $\rho = \ket{\Psi^+}\bra{\Psi^+}$, and we prove (\appref{app.sdef}) that to first order in perturbations away from the ideal state, fixing the angles gives the correct description of the decrease in $S$. As an alternative to $S$ one might calculate the fidelity of the postselected final state with respect to the ideal entangled state of two dual-rail qubits. The main goal of our analysis will be to determine the dependence of $S$ on the physical parameters in the setup.

The rate is determined by the probabilities for successful entanglement generation, connection, and postselection, denoted by $q_0$, $q_i$ where $i=1\dots n$, and $q_{ps}$ respectively. The average time $t_n$, it takes to create an entangled pair of length $L$, obeys the equation
\begin{equation}
\label{eq.avgtimereceq1}
t_{n+1} = q_{n+1}^{-1} (\tau 2^n  + t_n')
\end{equation}
where $t_n'$ is the average time it takes to create two neighbouring entangled pairs of length $L$ and $\tau=L_0/c$ is the classical communication time for an $L_0$-segment. The solution obtained by replacing $t_n'$ by $t_n$ in the above equation, i.e. by approximating the waiting time for two pairs by that of a single pair, considerably overestimates the rate. However, it turns out that a much better approximation is found by moving this assumption one step lower, i.e. by instead taking the recurrence for $t_n'$ to be
\begin{equation}
\label{eq.avgtimereceq2}
t_{n+1}' = \nu_{n+1} (\tau 2^n  + t_n') .
\end{equation}
Here $\nu_{n}$ is the average number of tries needed for two independent binomial events (i.e. generation/connection), each with probability $q_n$, to both succeed
\begin{equation}
\label{eq.avgnotries}
\nu_n = \frac{3-2q_n}{(2-q_n)q_n} .
\end{equation}
and the recurrence has solution
\begin{align}
t_n' & = \tau(2^{n-1}\nu_n + \dots + 2^0\nu_n \dots \nu_1 + \nu_n \dots \nu_1\nu_0)  \notag \\
     & \approx \tau \nu_n \dots \nu_1\nu_0 ,
\label{eq.avgtimereceq2sol}      
\end{align}
where the last equality holds for $q_0 \ll n^{-1}$ since for any linear optical Bell measurement the success probability is at most $1/2$ and therefore
\begin{equation}
\label{eq.nuinequalities}
\frac{2^{n-1}\nu_n + \dots + \nu_n \dots \nu_1}{\nu_n \dots \nu_1\nu_0} \leq \frac{n}{\nu_0} \leq n q_0 .
\end{equation}
Since to obtain good entangled states it is necessary to keep the photon creation probability $r^2$ small, this condition is well fulfilled in practice. Taking into account the final postselection step, we get the following expression for the rate
\begin{align}
\label {eq.rate}
R & = t_n^{-1} = \tau^{-1} q_{ps} \nu_{n}^{-1} \dots \nu_0^{-1}  \notag \\
  & \approx \tau^{-1} \left(\frac{2}{3}\right)^{n+1} q_{ps} q_{n} \dots q_0 ,
\end{align}
which is equivalent to the expression used in Refs. \cite{simon,sangouard,sangouard2}. It also agrees with the empirical estimate found in Ref. \cite{jiang}. The last simplification is exact in the limit of small $q$ and deviates from the first line of \eqnref{eq.rate} by at most a factor $\sim 2.6$ for $n\leq 8$. We have found that numerical simulation of the rate for given $q_i$ shows good agreement with \eqnref{eq.rate}.

\section{Methods}
\label{sec.methods}

To get at the figures of merit, we apply both analytical and numerical methods. Our approach is to compute the two mode density matrix $\rho_n$ of the entangled pairs at each step of the protocol (i.e. as a function of $L=2^nL_0$). Knowing the state, we may then calculate $S$ and any other derived quantities. In the course of computing $\rho_n$ we also obtain the success probabilities $q_i$ for entanglement generation and connection, which determine the rate. Although the methods we have developed are valid for arbitrary photon numbers, we shall in practise always work with small photon numbers, so that $\rho_n$ may be described by a 4x4 or 9x9 matrix in the Fock basis.

Computing $\rho_n$ may be broken into two principal steps. First, computing the state $\rho_0$ from entanglement generation (\figref{fig.entgenandcon} a) and second, computing $\rho_n$ for $n>0$ by iterating the connection process (\figref{fig.entgenandcon} b).

To deal with these two tasks, in particular the second one for which the Bogoliubov transformation depends on the atomic memory, we have developed a framework for calculating the output state from an arbitrary Bogoliubov transformation followed by projective measurements, given the input state. Our method, described in \appref{app.genfct}, is based on a \textit{generating function} $\gF$. This function takes two variables for each input and each output mode, and is defined such that its derivatives evaluated at zero form a matrix transforming the input to the output state, e.g., for a single input and output mode,
\begin{align}
\label{eq.genfct}
& \bra{i} \rho_{out} \ket{j} = \sum_{k,l} M_{ijkl}  \times \bra{k} \rho_{in} \ket{l}, \\
& M_{ijkl}                   = \left[\frac{1}{\sqrt{i!j!k!l!}}\frac{\partial^{k}}{\partial \alpha^{k}}\frac{\partial^{l}}{\partial \beta^{l}} \frac{\partial^{i}}{\partial \gamma^{i}} \frac{\partial^{j}}{\partial \delta^{j}} \gF (\alpha,\beta,\gamma,\delta)\right]_{\mathbf{0}} .
\end{align}
For any given Bogoliubov transformation and set of projection operators, we can compute $\gF$ and from $\gF$ we find $\rho_{out}$ for any given $\rho_{in}$.

In addition to the generating function, we also make use of \textit{mode reduction}. The Bogoliubov transformation for a full state transfer (i.e. storage and subsequent retrieval) of a light mode through a realistic atomic memory often involves many auxiliary modes in addition to the input and output modes. However, in \appref{app.moderedux} we show that the number of modes can always be reduced to three. The most general transformation for the state transfer becomes
\begin{equation}
\label{eq.storeandret}
\ah_1' = b_1 \ah_1 + c_1 \ah_1^\dagger + b_2 \ah_2 + c_2 \ah_2^\dagger + c_3 \ah_3^\dagger ,
\end{equation}
where unitarity requires that
\begin{equation}
\label{eq.unitarityreq}
|b_1|^2 + |b_2|^2 - |c_1|^2 - |c_2|^2 - |c_3|^2 = 1.
\end{equation}
Here the annihilation operator $\ah_1$ is for the input mode, and we have denoted the retrieved mode by a prime. The parameters $b_1,b_2,c_1,c_2,c_3$ are determined by the physical properties of the memory, and the state transfer is perfect when $b_1=1$ and only $b_1$ is non-zero. Additional terms are due to memory imperfections. We note that with the proper choice of phases, all but one of the parameters in \eqnref{eq.storeandret} may be assumed real. See \appref{app.moderedux} for details.

\begin{figure}
\centering
\includegraphics[width=.45\textwidth]{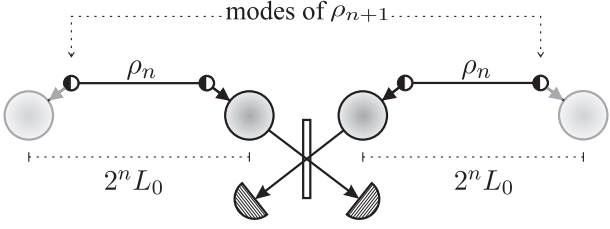}
\caption{Entanglement connection with no transmission losses. Two pairs of entangled light modes, each in state $\rho_n$ extending over a distance $2^n L_0$, are connected. Detection of a single click heralds successful connection and leads to entanglement between the far-away ends of the pairs, leaving them in the state $\rho_{n+1}$ extending a distance $2^{n+1} L_0$. A full state transfer through the memories is included in the connection step.}\label{fig.entconexample}. 
\end{figure}

As an example of how we apply mode reduction and the generating function, consider now a very idealised repeater setup, where the errors are small and due solely to memory imperfections. We neglect transmission losses and we assume the entanglement generation to produce a perfect Bell-state, $\rho_0 = \ket{\Psi^+}\bra{\Psi^+}$. The parameters of the memories are for now assumed to fulfill
\begin{equation}
\label{eq.goodmemcondition}
b_2,c_2,c_3 = 0, \qquad b_1^2 - 1 = c_1^2 \ll 1
\end{equation}
with $b_1,c_1$ real. Using these assumptions we can compute the generating function corresponding to entanglement connection and from the generating function we can find $\rho_n$. It is convenient to include the full state transfer through the memories (light to atoms to light) in the connection step such that $\rho_n$ denotes the state of two entangled light modes after $n$ connection steps. The connection then proceeds as shown in \figref{fig.entconexample}. The connecting ends of two neighbouring pairs of entangled light modes are stored in two memories, retrieved, mixed on a balanced beam splitter and then measured by non-counting photodetectors, with a single click heralding successful connection. 
The Bogoliubov transformation corresponding to connection may be found using \eqnref{eq.storeandret} and \eqnref{eq.goodmemcondition} and $\gF$ can then be computed from $\eqnref{eq.generalgenfct2}$. Using $\gF$ we find $\rho_n$ for the first few steps $n=1,2,...$, at each step expanding it to second order in $c_1^2$. Based on the results we come up with the following anzats:
\begin{widetext}
\begin{equation}
\label{eq.anzats}
\rho_n = \left(
\begin{array}{cccc}
1 - 2f_n + (2 f_n - 1 + 2 g_n) c_1^2 & 0 & 0 & (1 - 2f_n) c_1 \\
0 & f_n - g_n c_1^2 & f_n - g_n c_1^2 & 0 \\
0 & f_n - g_n c_1^2 & f_n - g_n c_1^2 & 0 \\
(1 - 2f_n) c_1 & 0 & 0 & (1 - 2f_n) c_1^2
\end{array}
\right),
\end{equation}
\end{widetext}
where $\rho_n$ is given in the Fock state basis, and $f_n,g_n$ are unknown, real valued functions to be determined. Our anzats will be confirmed, if the form \eqnref{eq.anzats} is preserved under entanglement connection (\figref{fig.entconexample}) and the resulting recursion equations for $f_n,g_n$ can be uniquely solved. Connecting two copies of \eqnref{eq.anzats} and expanding to second order in $c_1$, we find that the form is indeed preserved if
\begin{equation}
\label{eq.recexequations}
\begin{split}
f_{n+1} & = \frac{f_n}{2 - f_n}, \\
g_{n+1} & = \frac{4f_n(4 + g_n) + 11f_n^3 - 20f_n^2 - 4}{2f_n(f_n-2)^2}. 
\end{split}
\end{equation}
For $\rho_0$ to take the correct form, we must have $f_0 = \frac{1}{2}$, $g_0=0$, and these equations then yield the solution (see \appref{app.recexmath})
\begin{equation}
\label{eq.recexsolution}
\begin{split}
f_n & = \frac{1}{2^n + 1}, \\
g_n & = \frac{- 2^{1 + 3n} + 3\cdot2^{1 + 2n} + 5\cdot2^{n} - 9}{3\cdot2(2^n+1)^2}.
\end{split}
\end{equation}
Using $L/L_0=2^n$ we arrive at the following result for the conditional Bell parameter:
\begin{equation}
\label{eq.recexfidel}
S = 2\sqrt{2}(1 - (L/L_0 - 1)^2 c_1^2)
\end{equation}
to second order in $c_1$.

The above example illustrates the approach we take in deriving analytical results for the Bell parameter: Except for transmission losses in the detector arms, which may be considerable, errors are treated perturbatively and independently. First $\rho_0$ is computed, and then an anzats for $\rho_n$ to the desired order in the error is found, leading to recurrence equations which are solved with initial conditions given by $\rho_0$. We treat the following errors perturbatively (in each case keeping transmission losses finite): \textit{finite initial squeezing, dark counts in entanglement generation, memory imperfections}, and \textit{dark counts in entanglement connection}. The results for the conditional Bell parameter in each case are presented in the next section. 

To verify our analytical results, we have also performed numerical simulations, where it is not necessary to treat the errors perturbatively or independently. We have computed the generating functions for the Bogoliubov transformations corresponding to \figref{fig.entgenandcon} (a), (b) and made use of these functions to numerically compute $\rho_n$ for various values of the losses, initial squeezing and dark count rates. Numerical results for repeaters using two specific atomic memories are presented in section \secref{sec.num} and comparisons are made to the analytical approximations.

The success probabilities $q_i$ in the rate are found by inserting the Bogoliubov transformations corresponding to entanglement generation or connection in \eqnref{eq.generaloutputstate} and taking the trace. In this way one may in principle derive an expression for $R$ valid for a repeater based on a general memory, as we have done for $S$. As we shall see below, however, the protocol only works well if the memories are close to
being ideal apart from losses. We therefore only consider the rate for repeaters where losses are the sole errors since other imperfections will only
slightly perturb the results.

\section{Results}
\label{sec.results}

In the first section below we present our analytical results for the conditional Bell parameter in the presence of errors, and in the following section the analytical approximations are compared to numerical simulations of repeaters based on specific atomic quantum memories.

\subsection{Analytical results}
\label{sec.anal}

In the following subsections, we present results for $S(L/L_0)$ obtained by perturbation for each error source separately. Afterwards, we deal with cross terms between the perturbations. Arbitrary transmission losses are allowed in all cases and all results are given for both number resolving and non-number resolving photodetectors.

\begin{figure}
\centering
\includegraphics[width=.47\textwidth]{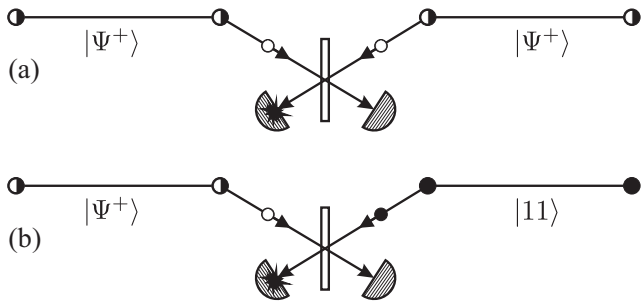}
\caption{Creation and propagation of errors. Filled circles denote excitations, empty circles denote vacuum. \textbf{(a)} During connection of two segments in $\ket{\Psi^+}$, vacuum is read out but the connection is accepted due to a detector dark count leaving the remaining modes in the separable state $\ket{11}$. \textbf{(b)} Once an error has occurred, it may propagate. Connecting $\ket{11}$ with and ideal entangled state and requiring a single click leads to $\ket{11}$ in the output (if no loss occurs). A vacuum state of the form $\ket{00}$ propagates in the same fashion.}\label{fig.conerrors}
\end{figure}

To get an intuitive idea of the nature of the errors, notice that the conditional Bell parameter $S$ will decrease whenever an erroneous event during generation or connection, such as a dark count or memory imperfection, can lead to a non-vacuum, separable output. That is, errors occur whenever superfluous excitations are introduced into the system. \figref{fig.conerrors}a shows how this may be caused by a connection dark count. Once an error has occurred, it propagates through the protocol as shown in \figref{fig.conerrors}b. Below we use such considerations to justify the scaling of $S$.

\subsubsection*{Imperfect memories and connection dark counts}

Dark counts during entanglement connection can be treated as a memory imperfection by considering the virtual PDC and beam splitter introduced in \figref{fig.entgenandcon} (b) to be a part of the memory protocol, and therefore we treat these two error sources simultaneously. An expression for the combined Bogoliubov transformation including dark counts is derived in \appref{app.moderedux}.

We take the Bogoliubov transformation for a full state transfer through the atomic memories including dark counts to be \eqnref{eq.dcfinalcoeff} and neglect all other error sources except transmission losses. The photon loss probability in entanglement connection is taken to be $p_{con}$. We treat each parameter in the transformation separately, and proceed as in the example given in \secref{sec.methods}. One can check that to leading order, this perturbation contains no cross-terms, and hence we may add the errors in $S$ deriving from each parameter. For the case without photon counting, we find
\begin{align}
\label{eq.fidelmemimperf}
\frac{S}{2\sqrt{2}} = 1 & - (L/L_0 -1)^2 (1+p_{con})^2 |c_1|^2 \\
                        & - 4 (L/L_0)^2 (1+p_{con})(|c_2|^2 + |c_3|^2 + \frac{\bar{n}_{dc}}{1-p_{con}}) \notag
\end{align}
and for the case with photon counting
\begin{align}
\label{eq.fidelmemimperfc}
\frac{S}{2\sqrt{2}} = 1 & - (L/L_0)^2 p_{con}^2 |c_1|^2 \\
                        & - 8 (L/L_0)^2 p_{con} (|c_2|^2 + |c_3|^2 + \frac{\bar{n}_{dc}}{1-p_{con}}) . \notag
\end{align}
For the perturbation to be valid, the error in $S$ must be less than 1, and so we require $c_1,c_2,c_3 \ll 1$ and $\bar{n}_{dc} \ll 1-p_{con}$. In addition, here we consider the large $L$ limit $L_0 \ll L$ and give only the leading order in $L$. However, we have verified that the expressions are a good approximation to the exact analytical result for $S$ (which could not be put on a closed form) also for $L \sim L_0$, as long as the perturbative condition is fulfilled.

There are several things to notice about the results \eqnref{eq.fidelmemimperf} and \eqnref{eq.fidelmemimperfc}. First, note that the conditional Bell parameter is independent of the parameter $b_2$. This is because $b_2$ corresponds to a plain loss. If $b_2$ is the only imperfection, the transformation \eqnref{eq.storeandret} is passive and hence $b_2$ leads to an increase of the vacuum component of $\rho_n$, which does not influence the conditional Bell parameter. For the same reason, there is no term in $S$ depending only on the transmission loss $p_{con}$. Second, note that the errors are suppressed for vanishing $p_{con}$, when the photons are counted, but persist for vanishing $p_{con}$, if they are not counted. This, and also the scaling of the error with $L$ and $p_{con}$, may be motivated by the following simple picture. 

In the case of only $c_1$ non-zero the unitarity condition \eqnref{eq.unitarityreq} becomes $|b_1|^2 - |c_1|^2 = 1$, and it follows that the effect of non-zero $c_1$ is a one-mode squeezing of the input mode. Similarly the effect of non-zero $c_2$ or $c_3$ is a two-mode squeezing of the input mode with an auxiliary mode. Letting $\rho = \rho_n \otimes \rho_n$ the memories take
\begin{equation}
\rho \rightarrow S_{L}(c_1)S_{R}(c_1) \rho S_{L}^\dagger(c_1)S_{R}^\dagger(c_1)
\end{equation}
or
\begin{equation}
\rho \rightarrow \text{Tr}_{lr}\left[ S_{Ll}(c_2)S_{Rr}(c_2) \rho S_{Ll}^\dagger(c_2)S_{Rr}^\dagger(c_2) \right]
\end{equation}
where (referring to \figref{fig.entconexample}) $L,R$ are the measured modes, $l,r$ are two auxiliary modes and $S_i(c)=\exp(c \ah_i^{\dagger 2} - c^* \ah_i^2)$ and $S_{ij}(c)=\exp(c \ah_i^\dagger \ah_j^\dagger - c^* \ah_i \ah_j)$ are the single and two-mode squeezing operators. From these expressions we see, that $c_1$ errors introduce photon pairs into the measured modes, while $c_2$ and $c_3$ errors introduce single photons. To lowest order in the $c$'s the amplitudes for errors to occur are $c_1,c_2,c_3$ and hence the error in $S$ scales with $|c_1|^2,|c_2|^2,|c_3|^2$. %
Now, the post-selection implied by $S$ (see \figref{fig.bellparamdef}) requires that any superfluous photons be removed before the post-selection stage. In the case of photon counting detectors, this can happen only through photon loss, and therefore the $c_1$ and $c_2,c_3$ error terms scale with $p_{con}^2$ and $p_{con}$ respectively. In the case of non-resolving detectors, in addition to loss, superfluous photons can be removed when multiple photons are incident on the same detector producing only a single click. This implies that errors persists for vanishing loss, and is apparent by the replacement $p_{con} \rightarrow (1+p_{con})/2$ from \eqnref{eq.fidelmemimperf} to \eqnref{eq.fidelmemimperfc}. %
To understand the scaling with $L$, note that there are $L/L_0-1$ connection attempts in total. There are therefore $~L/L_0$ ways for a photon to get lost and $~L/L_0$ ways for an extra photon to be introduced. In a $c_2$ or $c_3$ error one extra photon is introduced and thus one photon must get lost, whereas in a $c_1$ error two photons are added and two photon must get lost. However one of the added photons must get lost in the connection attempt in which it was created, since successful connection requires exactly one detector click, and the scaling for both types of error is therefore $L^2$. %
Since dark counts can be treated by mixing of the signal mode with a two mode squeezed state (see \figref{fig.entgenandcon}) this error term scales the same way as the $c_3$ term. This is also apparent from the Bogoliubov transformation \eqnref{eq.dcfinalcoeff}. Note that as $p_{con} \rightarrow 1$, the probability that a given click is a dark count approaches 1 and the dark-count error term diverges.


\subsubsection*{Finite initial squeezing}

Ideally, each entangled state consists of one excitation shared between two modes, and as seen above the introduction of additional excitations is a cause for errors. For this reason, when non-counting detectors are used or loss is present, the squeezing parameter of the state \eqnref{eq.twomodesqueeze} used for entanglement generation must be small and ideal Bell-type entanglement is only achieved in the limit of negligible squeezing $r \rightarrow 0$. Of course this ideal limit for the Bell parameter is the worst-case limit for the rate, as the success probability for entanglement generation is proportional to the probability $r^2$ of creating a photon pair in the PDC's. Thus, there is a trade-off between the conditional Bell parameter and the rate, and we are required to keep $r$ finite. Here we examine the effect of finite initial squeezing on $S$ for the final entangled state $\rho_n$.

The state $\rho_0$ produced by the setup \figref{fig.entgenandcon} (a) can be calculated either directly in the language of mode operators starting from the state \eqnref{eq.twomodesqueeze} or by computing the generating function of the entire setup including the PDC's and assuming vacuum input. We have found $\rho_n$ to second order in $r$, neglecting dark counts but making no restriction on the loss parameters $p_{gen},p_{con}$. Going to second order in $r$ is equivalent to taking a maximal photon number of two, and hence $\rho_0$ is described by a 9x9 matrix in this case. Knowing $\rho_0$ we proceed as in \secref{sec.methods} to find $\rho_n$, assuming perfect entanglement connection. The results for the conditional Bell parameter are
\begin{equation}
\label{eq.fidelfinsqueeze}
\frac{S}{2\sqrt{2}} = 1 - 8 (L/L_0)^2\, \frac{1+p_{gen}}{2} \frac{1+p_{con}}{2}\, r^2    
\end{equation}
without photon counting, and
\begin{equation}
\label{eq.fidelfinsqueezec}
\frac{S}{2\sqrt{2}} = 1 - 8 (L/L_0)^2\, p_{gen}\, p_{con}\, r^2       
\end{equation}
with photon counting.

The factor of $r^2$ indicates that, similar to what we saw previously, the decay of $S$ is caused by superfluous excitations in the repeater, in this case coming from the PDC's. The extra photons come in pairs, with one in the detector and one in the memory arm of \figref{fig.entgenandcon} (a), and hence the situation is analogous to that of $c_1$-type errors above. Two photons must get lost before the post-selection stage, and one of these must get lost in the generation attempt in which extra photons were introduced, since a single detector click is required for successful generation. With photon counting this type of error is suppressed when there is no loss in generation ($p_{gen} \rightarrow 0$) or connection ($p_{con} \rightarrow 0$), but persist for vanishing loss if the photons are not counted, and the error terms scales quadratically in $L$.

\subsubsection*{Generation dark counts}

Last, we consider the effect of dark counts in the entanglement generation, but neglect errors of second order or higher in $r$. Again, the generated state can be derived by means of the generating function for the setup \figref{fig.entgenandcon} (a). Having derived $\rho_0$ we proceed to find $\rho_n$ and the conditional Bell parameter assuming perfect entanglement connection. To lowest order in the dark count probability $\ndc$ we find, without photon counting
\begin{equation}
\label{eq.fidelgendc}
\frac{S}{2\sqrt{2}} = 1 - 4 (L/L_0)^2 \frac{1+p_{gen}}{1-p_{gen}} \bar{n}_{dc} ,
\end{equation}
and with counting
\begin{equation}
\label{eq.fidelgendcc}
\frac{S}{2\sqrt{2}} = 1 - 8 (L/L_0)^2\frac{p_{gen}}{1-p_{gen}} \bar{n}_{dc} .
\end{equation}

Again, notice that the error in $S$ is suppressed for $p_{gen} \rightarrow 0$ in the counting case but not in the non-counting case. Because a dark count alone results in the generation of a vacuum state, it must be combined with a double excitation in some segment to lead to an error in $S$ which explains the quadratic scaling of the error. Despite the fact that a double excitation is needed, $r^2$ does not show up in eqns. \eqnref{eq.fidelgendc}, \eqnref{eq.fidelgendcc}, because although the probability for a double excitation is of order $r^4$, there is no need to create an excitation in the segment where the dark count occurs, and hence the total number of generated photons is unchanged with respect to the ideal case.


\subsubsection*{Perturbation cross terms}

So far we have considered each error source independently, and we have found perturbatively their effect on the conditional Bell parameter. However, we have not addressed the possibility for cross terms in the perturbation, when errors of different type are present simultaneously, as will always be the case experimentally. Indeed, cross terms do appear and may have a severe effect on $S$ for certain values of the parameters. Here we identify the regime where cross terms can be safely neglected.

We find that the significant cross terms are those arising from the combination of a generation dark count with a memory imperfection or a connection dark count. To see this note, that a generation dark count results in a vacuum state, and connecting this with a Bell state in the presence of a memory imperfection or a generation dark count may result in the separable state $\ket{01}$, which leads to an error in $S$. Since the event requires a generation dark count and an error in connection, the error term must be proportional to $\epsilon \, \ndc$, with $\epsilon = |c_1|^2,|c_2|^2,|c_3|^2, \text{ or } \ndc$. However, compared with the errors considered in previous sections we now require generation of only $L/L_0 -1$ photons rather than $L/L_0$, and therefore this error term is also enhanced by a factor of $1/r^2$, so that the total order of magnitude of the term is $\epsilon \, \ndc/r^2$. Relative to the error terms considered previously, there is a factor of $\ndc/r^2$ which amounts to enhancement if the dark count probability is higher than the probability to generate a photon, and to suppression in the reverse case. On the other hand, the cross term involving a generation dark count and an additional generated photon, or the cross terms not involving generation dark counts must have $L/L_0$ generated photons. They are therefore of order $\epsilon^2$ and can be safely neglected in both cases.

We conclude that the analytical results derived in the previous sections provide a full description of the conditional Bell parameter whenever the production rate for photon pairs in the parametric down conversion is significantly higher than the dark count rate of the detectors. Since the errors due to finite initial squeezing and dark counts scale in roughly the same manner according to Eqns. \eqnref{eq.fidelmemimperf}, \eqnref{eq.fidelmemimperfc}, \eqnref{eq.fidelfinsqueeze}-\eqnref{eq.fidelgendcc}, it will be advantageous to make $r^2$ at least comparable with $\ndc$ to increase the rate. We are then only making a minor error by neglecting cross terms.

\subsection{Application to specific memories}
\label{sec.num}

We now verify the analytical results of the previous section by comparing with numerical simulations. The results for dark count and initial squeezing errors are independent of the memories and hence apply to any DLCZ repeater with the architecture presented above. We remind the reader that this includes the original DLCZ protocol since the interaction used by DLCZ for entanglement generation is effectively a two-mode squeezing and hence equivalent to our PDC's \cite{dlcz}. For the memory results we perform simulations using two specific ensemble based quantum memories. One proposed by Muschik and Hammerer \cite{muschik}, which we denote the two-pass memory, and one proposed by Julsgaard et al. \cite{julsgaard}, here denoted the one-pass memory. 

As explained in \secref{sec.methods} our numerical results are obtained by computing $\rho_n$ by means of the generating function for specific values of the parameters of the system. In doing so, there is no need for treating errors perturbatively but we do have to restrict the dimension of $\rho_n$, because it is not practical to work with large matrices. Since $\rho_n$ is given in the Fock state basis, this effectively means restricting the maximal number of excitations in the entangled states. In our simulations we take the maximal excitation number to be 2 so that $\rho_n$ is 9x9. This implies that our numerical results can be considered exact as long as the probability of creating three or more excitations is negligible, however for the repeater to work reliably beyond a few connection steps $r^2 \ll 1$ is required, so this condition is well fulfilled in practice. 

\begin{figure}
\centering
\includegraphics[width=.47\textwidth]{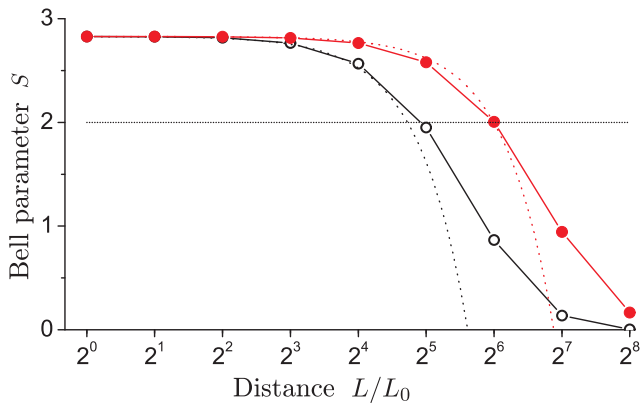}
\caption{The effect of finite initial squeezing. $S(L)$ is plotted for $r=10^{-2}$ and transmission losses of $p_{gen}=0.9$, $p_{con}=0.1$, in the case of counting (dots) and non-counting (circles) detectors. Dashed lines show the analytical approximations.}\label{fig.sqerr}
\end{figure}

\begin{figure}
\centering
\includegraphics[width=.47\textwidth]{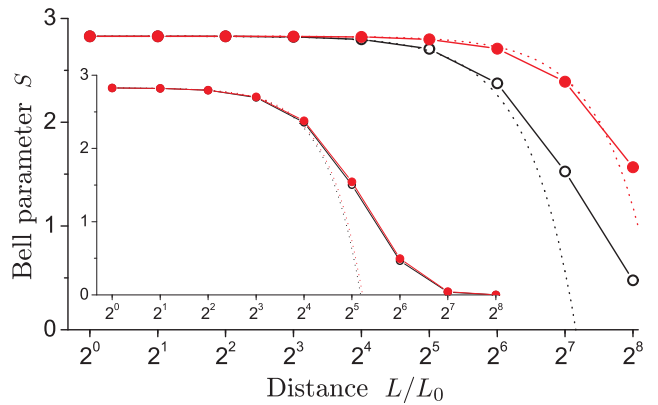}
\caption{Effect of detector dark counts in entanglement connection. $S(L)$ is plotted for $\ndc=10^{-4}$ and transmission losses of $p_{gen}=0.9$, $p_{con}=0.1$, for counting (dots) and non-counting (circles) detectors. Dashed lines are analytical results. The inset shows the effect of dark counts in entanglement generation, for the same parameter values.}\label{fig.dcerr}
\end{figure}

Results for the effects of finite initial squeezing and of dark counts on the conditional Bell parameter are displayed in Figs. \ref{fig.sqerr},\ref{fig.dcerr}. For each type of error we display the analytical conditional Bell parameter from the previous section as well as the numerical simulation. Note that in all cases, $S$ is well described by the analytical approximations for distances $L$ where $S$ is well above the classical threshold of 2. The excellent agreement confirms the analytical results of \secref{sec.anal}. The perturbative approximation breaks down at the distance where $S$ drop below the classical threshold of 2, i.e. the maximal distance over which communication is possible with the given losses, since at this point the error in $S$ is no longer small. 

We can estimate the maximal distance imposed by dark counts or PDC squeezing by setting $S=2$ in \eqnref{eq.fidelfinsqueeze} and \eqnref{eq.fidelgendc} to get
\begin{align}
(L/L_0)^2  & = \frac{2-\sqrt{2}}{4(1+p_{gen})(1+p_{con})} r^{-2}           \label{eq.sqrestrict}     \\    
(L/L_0)^2  & = \frac{2-\sqrt{2}}{8} \frac{1-p_{gen}}{1+p_{gen}} \ndc^{-1}  \label{eq.rndcrestrict}         
\end{align}
in the non-counting case. Small $r$ are not hard to obtain experimentally, but the dark count number $\ndc$ is dictated by the detector dark count rate and the pulse duration $\tau$. With microsecond pulses, an optimistic 1Hz dark count rate, and 90\% photon loss in generation, the maximal distance \eqnref{eq.rndcrestrict} is $L \sim 64L_0$. Recent protocols offer better tolerance for dark counts and other multi-photon errors, see e.g. Ref. \cite{jiang}.

\begin{figure}
\centering
\includegraphics[width=.47\textwidth]{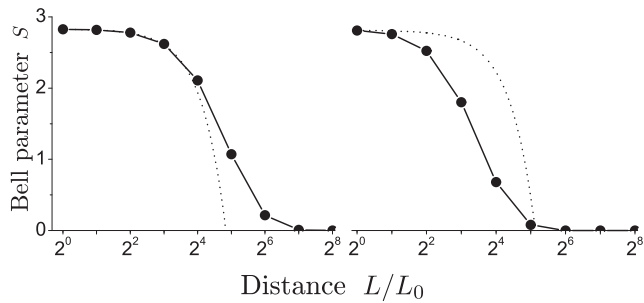}
\caption{Significance of perturbation cross terms. $S(L)$ is plotted with dark counts in all detectors for $\ndc=10^{-5}$ in the two cases $r^2=10^{-4}$ (left )and $r^2=10^{-6}$ (right). When $\ndc>r^2$, a large discrepancy is observed between the numerical result and the analytical approximation omitting perturbation cross terms (dashed line).}\label{fig.crossterms}
\end{figure}

As mentioned in the previous section, the different errors introducing superfluous excitations in the repeater can only be treated independently in perturbation theory when the production rate in the PDC's is higher than the rate of spurious excitations. This is demonstrated in \figref{fig.crossterms}, where we plot $S$ in the two cases $\ndc < r^2$ and $\ndc > r^2$ when both generation and connection dark counts occur. We see that the analytical approximation obtained by adding the error terms in \eqnref{eq.fidelmemimperf} and \eqnref{eq.fidelgendc} is valid only in the former case

\begin{figure}
\centering
\includegraphics[width=.35\textwidth]{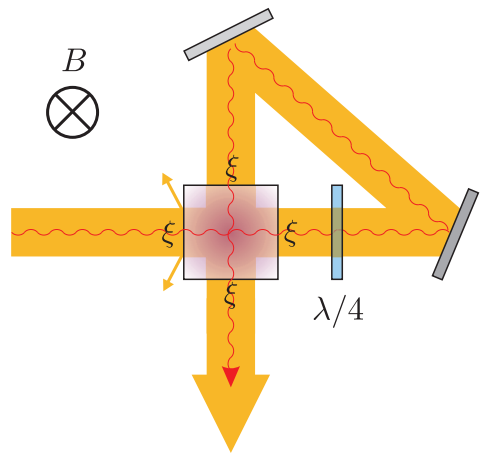}
\caption{Setup for the two-pass memory. An atomic ensemble is placed in a magnetic field, with the spin of the ensemble strongly polarised along the field. The atoms are traversed twice in orthogonal directions in the plane perpendicular to the field by a light beam. As a result, the light polarisation state can be stored in or retrieved from the atomic spin.}\label{fig.twopasssetup}
\end{figure}

Next we turn to specific quantum memories. The two-pass memory is depicted in \figref{fig.twopasssetup}. It consists of an ensemble of atoms contained in a glass cell at room temperature and traversed twice in orthogonal directions by a light beam. Both light and atomic spin are strongly polarised, such that the Holstein-Primakoff approximation may be applied and each system can be described by a harmonic oscillator. The interaction Hamiltonian is $\hat{x}_A\hat{x}_L$ in one pass of the light pulse and $\hat{p}_A\hat{p}_L$ in the other, where subscripts A and L refer to atoms and light, and the overall interaction Hamiltonian becomes $\hat{x}_A\hat{x}_L + \hat{p}_A\hat{p}_L \sim \ah_A^\dagger\ah_L + \ah_A\ah_L^\dagger$. As a result of this interaction, the polarisation state of the light and the state of the collective atomic spin are swapped, and this process is governed by the light-atom interaction strength $\kappa$. In the absence of losses a full state transfer through the memory may be described by the Bogoliubov transformation (see Ref. \cite{muschik} for details)
\begin{equation}
\label{eq.fullstatetransf}
\ah_L' = (e^{-\kappa^2} - 1) \ah_L - e^{-\kappa^2/2}\sqrt{1-e^{-\kappa^2}} \ah_A + e^{-\kappa^2/2} \ah_L^{ret} .
\end{equation}
where $\ah_L,\ah_L'$ are the stored and retrieved light modes, $\ah_A$ is a collective atomic mode and $\ah_L^{ret}$ is the input mode of the retrieval light pulse. From \eqnref{eq.fullstatetransf} it is apparent that the memory is perfect when $\kappa$ is large. However, in experiment $\kappa$ is restricted by spontaneous emission, and in the following examples we use an optimistic value of $\kappa = 2$. In the lossless case \eqnref{eq.fullstatetransf} is a passive transformation and hence does not introduce any error in $S$ (c.f. \eqnref{eq.storeandret},\eqnref{eq.fidelmemimperf},\eqnref{eq.fidelmemimperfc}). Only the rate is affected by $\kappa$.

\begin{figure}
\centering
\includegraphics[width=.47\textwidth]{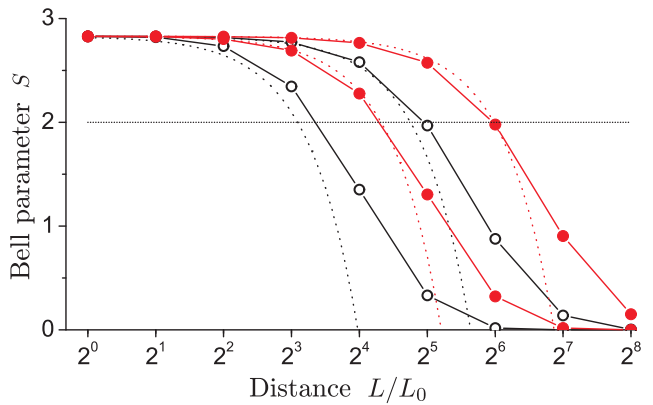}
\caption{Effect of two-pass memory imperfections. $S(L)$ is plotted for $\xi=10^{-4},10^{-3}$ and transmission losses of $p_{gen}=0.9$, $p_{con}=0.1$, for counting (dots) and non-counting (circles) detectors. Dashed lines are analytical results.}\label{fig.memerr}
\end{figure}

The important error parameter in the setup \figref{fig.twopasssetup} is the reflection coefficient $\xi$ for reflections at the cell walls. Reflections occurring between the two passes of the light pulse introduce an active part to the transformation \eqnref{eq.fullstatetransf} leading to non-zero $c_3$ in \eqnref{eq.storeandret}, and in fact we find $c_3^2 \approx 0.9\xi$ to lowest order in $\xi$ when $\kappa=2$ \cite{privcommuschik}. Putting $S=2$ in \eqnref{eq.fidelmemimperf} we have
\begin{equation}
\xi = \frac{2-\sqrt{2}}{0.9\cdot 8(1+p_{con})} (L/L_0)^{-2} .
\label{eq.xirestrict}
\end{equation}
Taking an optimistic value $\xi = 10^{-3}$ and no loss, this limits the communication distance to $L \sim 8L_0$. In \figref{fig.memerr} we plot $S$ for the two-pass memory together with our numerical approximation.

\begin{figure}
\centering
\includegraphics[width=.47\textwidth]{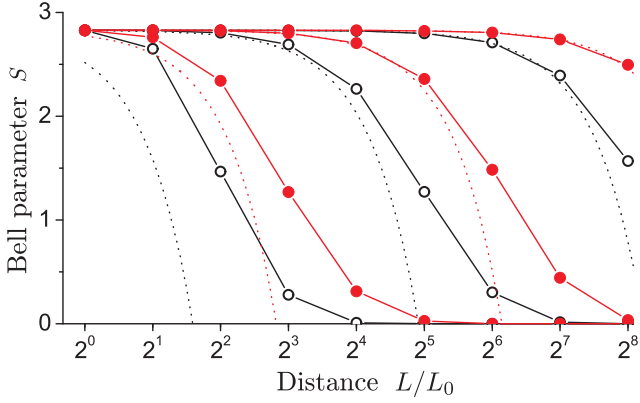}
\caption{Effect of one-pass memory imperfections. $S(L)$ is plotted for squeezing parameters (from the top down) $s=10^5,10^3,10$ and transmission losses of $p_{gen}=0.9$, $p_{con}=0.1$ for counting (dots) and non-counting (circles) detectors. Dashed lines are analytical results.}\label{fig.onepassmemerr}
\end{figure}

The one-pass memory has been demonstrated experimentally by Julsgaard et al. \cite{julsgaard}. The write-in setup is similar to the two-pass setup in \figref{fig.twopasssetup}, but instead of the polariser a detector is placed is placed in the beam and the light is measured after one pass through the atomic ensemble. A feedback is then supplied to the atoms via a magnetic pulse based on the measurement outcome. In order to see whether this memory is suitable for use in the repeater, we simply assume perfect readout. With this assumption the Bogoliubov transformation of a full state transfer becomes
\begin{equation}
\ah_L' = (1-\frac{\kappa g}{2})\ah_A + \frac{\kappa g}{2}\ah_A^\dagger + \frac{1}{2}(\kappa + g)\ah_L - \frac{1}{2}(\kappa - g)\ah_L^\dagger ,
\label{eq.onepasstransf}
\end{equation}
where $\kappa$ is the coupling strength and $g$ is the feedback gain. From this expression it is clear that the memory is never perfect, since for any non-zero choice of $\kappa$ and $g$ the output light contains some mixing in of the atomic mode. However, if the atomic mode is squeezed prior to storage, the noise introduced by the atoms can be suppressed. Assuming that the variance in the $X$-quadrature of the atomic mode is squeezed by a factor $s$, and putting $\kappa=g=1$ the Bogoliubov of a state transfer becomes
\begin{equation}
\ah_L' = \frac{\sqrt{s}}{2}(\ah_A + \ah_A^\dagger) + \ah_L .
\label{eq.onepassbogoliubov}
\end{equation}
This is on the form \eqnref{eq.storeandret} and it is now easy to read of the coefficients and plug into \eqnref{eq.fidelmemimperf}. Taking $S=2$ and solving for $s$ we get
\begin{equation}
s = \frac{1}{2} \frac{2-\sqrt{2}}{1+p_{con}} (L/L_0)^{-2} \approx - 30\,\text{dB}
\label{eq.squeezingrestrict}
\end{equation}
for $L=16L_0$ and no loss. This value of $s$ is far beyond what can be achieved experimentally at the moment \cite{julsgaard,kuzmich,geremia,fernholz}. E.g. in one recent experiment about $3\,\text{dB}$ of squeezing was reported \cite{fernholz}. We therefore conclude that the one-pass memory is not suitable for implementation of a DLCZ-type repeater. The performance for several values of the squeezing is shown in \figref{fig.onepassmemerr}

\begin{figure}
\centering
\includegraphics[width=.45\textwidth]{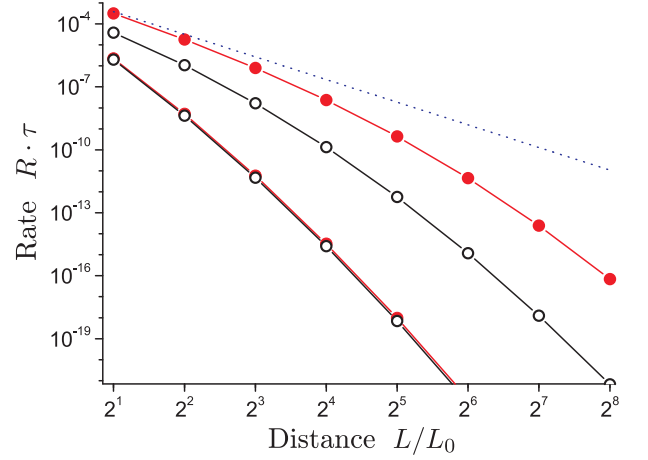}
\caption{The rate for a fixed imperfection in $S$ of 5\%. We plot the two cases of counting (circles) and non-counting (dots) detectors, for $p_{gen}=0.9$ and $p_{con}=0.1$ (upper curves), $p_{con}=0.9$ (lower curves). For reference we plot the ideal scaling $L^{-2-\log_2{3}}$ (dashed line).}\label{fig.rates}
\end{figure}

Finally, let us briefly consider the rate. We prove in \appref{app.dlczrate}, that the rate of the DLCZ-repeater in the absence of dark counts and memory imperfections is given by
\begin{equation}
\label{eq.translossrateclosednocount}
R = \frac{2}{3} r^2 (1-p_{gen}) (L/L_0)^{-\log_2{3/2}} R'
\end{equation}
with
\begin{equation}
\label{eq.rprimenocount}
R' = e^{\frac{1}{\ln 2} \text{Li}_2\left(\frac{p_{con}+1}{p_{con}-1} \sqrt{2} L/L_0\right) - \frac{1}{\ln 2} \text{Li}_2\left(\frac{p_{con}+1}{p_{con}-1} \sqrt{2}\right)} .
\end{equation}
in the non-counting case, and
\begin{equation}
\label{eq.translossrateclosed}
R = \frac{2}{3} r^2 (1-p_{gen}) (L/L_0)^{-\log_2{3}} R'
\end{equation}
with
\begin{equation}
\label{eq.rprime}
\begin{split}
R' = e^{\frac{1}{\ln 2} \text{Li}_2\left(\frac{p_{con}}{p_{con}-1} \sqrt{2}L/L_0\right) - \frac{1}{\ln 2} \text{Li}_2\left(\frac{p_{con}}{p_{con}-1} \sqrt{2}\right)}
\end{split}
\end{equation}
in the counting case. We can obtain the scaling of the rate for a fixed final imperfection in $S$ , by using \eqnref{eq.fidelfinsqueeze} or \eqnref{eq.fidelfinsqueezec} to determine the value of $r$ and inserting this value into the above expressions. \figref{fig.rates} shows the result. In the best case, when the connection loss is negligible, the scaling is $L^{-2-\log_2{3}}$. When connection losses are small, the rate is seen to be significantly enhanced by the use of counting detectors. This can be understood as a consequence of the vacuum component of the state $\rho_n$ growing faster with $n$ for non-counting detectors. Connecting two entangled pairs using non-counting detectors may lead to a vacuum state even in the absence of any losses. This is not the case for counting detectors. A fast growing vacuum component leads to a low connection probability, since the vacuum cannot contribute to the clicks required for successful connection. As losses increase, the probability for two photons to reach a detector simultaneously decreases, and hence the advantage of counting over non-counting detectors disappears, the two rates being equal in the limit of very high loss.  


\section{Conclusion}

We have presented a thorough analysis of quantum repeaters using general ensemble based memories in the DLCZ architecture. As a primary result of our analysis we have derived perturbative analytical expressions for the Bell parameter of the generated entangled states in terms of the distance and the memory parameters. Our results apply to repeaters based on any quantum memory which may be described by a Bogoliubov transformation and hence to any system for which the interaction is no more than quadratic in the mode operators. We have verified our analytical results by comparison to numerical simulations and found good agreement in the range where perturbation theory applies.

We are aware that many protocols improving upon the DLCZ repeater architecture have been put forward, which promise significantly better tolerance for multi-photon errors and better rates, and that hence the system analysed in this paper is not a likely candidate for experimental implementation \cite{chen,jiang,simon,sangouard,collins}. However, we have demonstrated how to analyse a repeater in terms of a very general quantum memory model, and our methods can in principle be extended to any system for which entanglement generation and creation are described by Bogoliubov transformations and single-photon detection. This is the case for the recent proposals Refs. \cite{chen,jiang,simon,sangouard,collins}.

\acknowledgments

We acknowledge K. Mølmer and J. Sherson for initial ideas on the generating function formalism and thank C. Muschik and K. Hammerer for sharing their calculations on the two-pass memory with us. We also acknowledge helpful discussions with M. Wubs and H. J. Kimble.

\appendix

\section{Bell parameter angles}
\label{app.sdef}

Here we prove that the angles in the definition of $S$ can be held fixed when calculating the decrease in $S$ in first order perturbation theory. Let $\mbox{\boldmath $\phi $}^0 = (\phi_1^0,\ldots,\phi_4^0)$ be the angles maximising $S(\ket{\Psi^+}\bra{\Psi^+})$ and let $\rho(\mathbf{x})$ be a perturbation away from the Bell state with $\rho(\mathbf{0})=\ket{\Psi^+}\bra{\Psi^+}$. Then the angles which maximise $S(\rho)$ will also be close to $\mbox{\boldmath $\phi $}^0$ and to first order in the perturbation, we can write
\begin{align}
S(\mathbf{x},\mbox{\boldmath $\phi $}) & = 2\sqrt{2} + \sum_{i=1}^k \frac{\partial S}{\partial x_i}\bigg|_{\mbox{\boldmath $0,\phi $}^0}\cdot x_i + \sum_{i=1}^4 \frac{\partial S}{\partial \phi_i}\bigg|_{\mbox{\boldmath $0,\phi $}^0}\cdot \Delta_{\phi_i}  \notag \\
 & = 2\sqrt{2} + \sum_{i=1}^k \frac{\partial S}{\partial x_i}\bigg|_{\mbox{\boldmath $0,\phi $}^0}\cdot x_i \notag \\
 & = S(\mathbf{x},\mbox{\boldmath $\phi $}^0) ,
\end{align}
where $\Delta_{\phi_i} = \phi_i - \phi_i^0$ and we have used that $\frac{\partial S}{\partial \phi_i}\big|_{\mbox{\boldmath $0,\phi $}^0}=0$. Thus we have proven that to first order in the perturbation, it is optimal to evaluate $S$ using the angles $\mbox{\boldmath $\phi $}^0$.

\section{The generating function}
\label{app.genfct}

We define the generating function in the general case, where our system $\mathcal{S}$ has arbitrarily many modes, some of which are output while the remaining modes are measured or traced out. Symbolically $\mathcal{S = OR}$, where $\mathcal{O}$ are the output and $\mathcal{R}$ the remaining modes. If the system is subject to a Bogoliubov transformation $U$ and a subsequent measurement with an outcome corresponding to projection operator $P$, then the unnormalised output state is
\begin{equation}
\label{eq.generaloutputstate}
\rho_\mathcal{O}^{out} = \text{Tr}_\mathcal{R}(P U\rho_\mathcal{S}^{in}U^{\dagger} P^\dagger) .
\end{equation}

To construct the generating function, first note that a Fock state in mode $i$ may be written
\begin{equation}
\label{eq.fockstateexp}
\ket{n}_i = \frac{1}{\sqrt{n!}} (\ah_i^\dagger)^n \ket{0}_i = \left[ \frac{1}{\sqrt{n!}} \frac{\partial^n }{\partial \alpha^n } e^{\alpha \ah_i^\dagger} \ket{0}_i \right]_{\alpha = 0} 
\end{equation}
where the parameter $\alpha$ can be chosen real and $\ket{0}_i$ denotes the vacuum state of mode $i$. The generating function is defined to be
\begin{widetext}
\begin{align}
\label{eq.generalgenfct1}
\gF(\alpha_i,\beta_i,\gamma_i,\delta_i) & = \text{Tr}_{\mathcal R}\left[
_{\mathcal O}\bra{0}
\left(\prod_o e^{\delta_o \ah_o}\right)
P U
\left(\prod_i e^{\beta_i \ah_i^\dagger}\right)
\vacr\vacl
\left(\prod_i e^{\alpha_i \ah_i}\right)
U^\dagger P^\dagger
\left(\prod_o e^{\gamma_o \ah_o^\dagger}\right)
\ket{0}_{\mathcal O}
\right] \\
\label{eq.generalgenfct2}
& = e^{\frac{1}{2}\Sigma} \times \vacl
\left(\prod_i D_i(-\alpha_i)\right)
U^\dagger P^\dagger
\left(\prod_o D_o(\gamma_o) |0\rangle_o\langle 0| D_o(-\delta_o)\right)
P U
\left(\prod_i D_i(\beta_i)\right)
\vacr 
\end{align}
\end{widetext}
where $i$ runs over all of $\mathcal S$ and $o$ runs over the output modes $\mathcal O$. The parameters $\alpha_i,\beta_i$ and $\gamma_i,\delta_i$ are real and correspond to the input and output modes respectively. Using \eqnref{eq.fockstateexp} and \eqnref{eq.generalgenfct1} we can generalise \eqnref{eq.genfct}, and we see that $\gF$ generates the output state as desired. The expression \eqnref{eq.generalgenfct2} with $\Sigma \equiv \sum_i(\alpha_i^2 + \beta_i^2)+\sum_o(\gamma_o^2 + \delta_o^2)$ follows from the circular property of the trace and the relation
\begin{equation}
\label{eq.fockstatedispop}
e^{\alpha \ah^\dagger} \ket{0} =  e^{\frac{1}{2}|\alpha|^2} D(\alpha) \ket{0} .
\end{equation}
which is a consequence of the disentangling theorem \cite{gerry}.

We now show how to obtain the function $\gF$ for given $U$ and $P$. Let the Bogoliubov transformation $U$ be given by
\begin{equation}
\label{eq.generalbogoliubov}
U^{\dagger} \ah_j U = \sum_i b_{ji} \, \ah_i + c_{ji} \, \ah_i^{\dagger}.
\end{equation}
To compute $\gF$ we make use of several properties of displacement operators. First, one may prove that the vacuum projection operator can be written as the integral
\begin{equation}
\label{eq.projbydisp}
|0\rangle\langle 0| = \int\frac{dpdx}{2\pi} e^{-(x^2+p^2)/4} D\left(\frac{x+ip}{\sqrt{2}}\right) .
\end{equation}
It follows from \eqnref{eq.fockstateexp} that the projection on any Fock state can be written in terms of derivatives of an integral over a product of displacement operators. Second, under $U$ the displacement operators transform as
\begin{equation}
\label{eq.disptransform}
U^\dagger D_j(\beta) U = \prod_i D_i(\beta b_{ji}^* - \beta^* c_{ji}) .
\end{equation}
And third, the product and vacuum expectation value of displacement operators are given by
\begin{align}
\label{eq.dispproduct}
D(\alpha)D(\beta) & = e^{i \text{Im}(\alpha \beta^*)} D(\alpha + \beta) , \\
\label{eq.vacdisp}
\bra{0} D(\alpha)\ket{0} & = e^{-\frac{1}{2}|\alpha|^2} .
\end{align}

Starting from \eqnref{eq.generalgenfct2}, the function $\gF$ is found in four steps. First, all projection operators in the expression are replaced by integrals of displacement operators by making use of \eqnref{eq.projbydisp}. Second, the Bogoliubov transformation is eliminated from the expression via \eqnref{eq.disptransform}. Third, the integrals are pulled outside the vacuum expectation which then contains only a product of displacement operators, and the expectation value is evaluated by using \eqnref{eq.dispproduct} and \eqnref{eq.vacdisp}. Last, the resulting Gaussian function is integrated and we obtain an analytic expression for $\gF$ involving only the $\alpha,\beta,\gamma,\delta$-variables and the parameters of $U$.

\subsubsection*{Projection operators}

The generating function as defined above can be computed for any measurement described by a projection in the Fock state basis. However, in this article we are interested only in measurements where a single or no click is observed. The measurement operator corresponding to the zero outcome (no click) is simply a projection on the vacuum state of the measured mode:
\begin{equation}
\label{eq.measopnoclick}
P_{dark} = \ket{0}\bra{0}.
\end{equation}
The measurement operator corresponding to a single click depends on the resolution properties of the detector. We work with two contrasting cases. For perfect single-photon counters, the operator is
\begin{align}
\label{eq.measopclickcount}
P_{light} & = \ket{1}\bra{1} \\
          & = \left[ \frac{\partial^2 }{\partial \alpha \partial \beta} e^{(|\alpha|^2 + |\beta|^2)/2} D(\alpha) \ket{0}\bra{0} D(\beta) \right]_{\alpha,\beta = 0} , \notag
\end{align}
where \eqnref{eq.fockstateexp} and \eqnref{eq.fockstatedispop} was used. For detectors which can only distinguish between presence and absence of light but give no information about the photon number, the operator becomes
\begin{equation}
\label{eq.measopclicknocount}
P_{light} = \mathbbm{1} - \ket{0}\bra{0} .
\end{equation}

\subsubsection*{Squeezed initial states}

In principle the generating function \eqnref{eq.generalgenfct1} allows for any input and output states. In practice, it is not convenient to work with high photon numbers, because the density matrices become large and the calculation of the elements corresponding to many-photon Fock states require derivatives of high orders (c.f. \eqnref{eq.genfct}). One might therefore expect that squeezed initial states would be treated only approximately, as they have non-zero overlap with all the Fock states. Fortunately, the squeezed states belong to a class of initial states which may be treated exactly: For any state, which can be written on the form $A\vacr$, where the operator $A$ generates a Bogoliubov transformation, we see from \eqnref{eq.generaloutputstate} that we may replace $U$ by $UA$ and take $\rho^{in}_\mathcal{S} = \vacr\vacl$. In particular, for the two-mode squeezed state \eqnref{eq.twomodesqueeze} in modes $i,j$, we let $U \rightarrow U S_{ij}(r)$ where $S_{ij}(r)$ is the usual two-mode squeezing operator. In this way the input is treated exactly, regardless of photon number.


\section{Mode reduction}
\label{app.moderedux}

In this appendix we show how to reduce the most general Bogoliubov transformation to three modes, and how to include dark counts in the memory Bogoliubov transformation. The most general Bogoliubov transformation for a full state transfer through a quantum memory takes the form
\begin{equation}
\label{eq.generalstatetrans}
\ah_1' = b_1 \ah_1 + c_1 \ah_1^\dagger + \sum_i \tilde{b}_i\hat{\tilde{a}}_i + \tilde{c}_i\hat{\tilde{a}}_i^\dagger
\end{equation}
where $\ah_1$ is the mode operator of the stored mode, and $\ah_1'$ that of the retrieved mode. Now, it is always possible to define new, independent mode operators $\ah_2,\ah_3$ by
\begin{equation}
\label{eq.newmodeops}
b_2 \ah_2 \equiv \sum_i \tilde{b}_i\hat{\tilde{a}}_i , \quad
c_2 \ah_2^\dagger + c_3 \ah_3^\dagger \equiv \sum_i \tilde{c}_i\hat{\tilde{a}}_i^\dagger ,
\end{equation}
where $b_2,c_2,c_3$ are complex coefficients, and hence we can simplify \eqnref{eq.generalstatetrans}:
\begin{equation}
\label{eq.reduxstatetrans}
\ah_1' = b_1 \ah_1 + c_1 \ah_1^\dagger + b_2 \ah_2 + c_2 \ah_2^\dagger + c_3 \ah_3^\dagger .
\end{equation}
We see that the new transformation involves only three modes.

The b,c coefficients are determined by \eqnref{eq.newmodeops} and the canonical commutator relations for $\ah_2,\ah_3$. There is some freedom in the choice of phases however. The phases of $b_1$, $c_1$ can be adjusted by simple phase shifts of the input and output modes. It can be seen, that in our repeater setup a phase shift on the output mode has no effect on the measurement outcomes when connecting pairs, and hence we may always assume that either $b_1$ or $c_1$ is real. It is not obvious that choosing both of them real corresponds to an optimal phase choice in terms of $S$, however we have checked numerically that a phase change of the input mode has negligible effect on $S$. Hence in this paper we take $b_1$ and $c_1$ to be real. Any complex phase on $b_2$ can be absorbed into the definition of $\ah_2$, and likewise the phase of $c_3$ can be absorbed in $\ah_3$. We may therefore assume $b_2$, $c_3$ to be real. Putting things together, we have
\begin{equation}
\label{eq.reduxphases}
b_1,b_2,c_1,c_3 \in \mathbb{R} , \qquad c_2 \in \mathbb{C} .
\end{equation}
And because $\ah_1'$ must preserve the canonical commutation relations
\begin{equation}
\label{eq.reducnormrelation}
b_1^2 + b_2^2 - c_1^2 - |c_2|^2 - c_3^2 = 1 .
\end{equation}

\subsubsection*{Including dark counts}

As an example of mode reduction, we consider dark counts in entanglement connection. They are treated by including a PDC and beam splitter after the memory readout, as shown in \figref{fig.entgenandcon} (b). Tracing out one mode of the PDC leaves a thermal state in the other mode, which is mixed with the signal from the memory. We now show how the memory transformation \eqnref{eq.reduxstatetrans} can be modified to take account of the dark counts. 

The new Bogoliubov transformation, including the additional modes $\ah_{s1},\ah_{s2}$, is given by $U_{BS} U_{mem} S_{s1s2}(r)$, where $U_{BS}$ is a beam splitter transformation, $U_{mem}$ is the transformation \eqnref{eq.reduxstatetrans} and we have included the squeezing as explained in \appref{app.genfct}. This implies
\begin{align}
\label{eq.dcnewbogoliubov}
\ah_1'' & = S_{s1s2}^\dagger(r)(\ah_1' \sqrt{1-p}  + \ah_{s1} \sqrt{p})S_{s1s2}(r) \notag \\
        & = \ah_1' \sqrt{1-p}  + \sqrt{p}\,(\ah_{s1} \cosh s - \ah_{s2}^\dagger \sinh s) .
\end{align}
Using \eqnref{eq.reduxstatetrans} and applying mode reduction, we can write
\begin{equation}
\label{eq.dcreduxstatetrans}
\ah_1'' = b_1' \ah_1 + c_1' \ah_1^\dagger + b_2' \ah_2' + c_2' \ah_2'^\dagger + c_3' \ah_3'^\dagger ,
\end{equation}
where
\begin{align}
\label{eq.dcmodedef}
b_2' \ah_2'                               & \equiv \sqrt{1-p}\, b_2 \ah_2 + \sqrt{p} \cosh(s) \, \ah_{s1} , \\
c_2' \ah_2'^\dagger + c_3' \ah_3'^\dagger & \equiv \sqrt{1-p}\,(c_2 \ah_2^\dagger + c_3 \ah_3^\dagger) - \sqrt{p} \sinh(s) \ah_{s2}^\dagger . \notag
\end{align}
Choosing $b1',b2',c3'$ real and positive and using the canonical commutators, we find the primed coefficients to be
\begin{align}
\label{eq.dcnewcoeff}
b_1' & = \sqrt{1-p}\, b_1 \notag \\
b_2' & = \sqrt{(1-p) b_2^2 + p \cosh(s)^2} \notag \\
c_1' & = \sqrt{1-p}\, c_1 \\
c_2' & = (1-p) b_2 c_2 / b_2' \notag \\
c_3' & = \sqrt{(1-p)(|c_2|^2 + c_3^2) + p \sinh(s)^2 - |c_2'|^2} . \notag
\end{align}

It remains to relate the parameters $p,s$ of the virtual optical elements to the physical dark count rate. One may prove, that the output state from one arm of the PDC, when the other arm is traced out, is a thermal state of mean photon number $\sinh(s)^2$. The average number of dark counts must equal the mean photon number at the detector due to the virtual PDC, and we therefore get $\bar{n}_{dc} = p\sinh(s)^2$. To obtain a final expression for the Bogoliubov transformation including dark counts we rewrite \eqnref{eq.dcnewcoeff} in terms of the physical parameter $\bar{n}_{dc}$, and since we are only interested in introducing dark counts but not photon loss to the memory output mode we let $p \rightarrow 0$ while keeping $\bar{n}_{dc}$ constant. The result is
\begin{align}
\label{eq.dcfinalcoeff}
b_1' & = b_1 \notag \\
b_2' & = \sqrt{b_2^2 + \bar{n}_{dc}} \notag \\
c_1' & = c_1 \\
c_2' & = b_2 c_2 / b_2' \notag \\
c_3' & = \sqrt{|c_2|^2 - |c_2'|^2 + c_3^2 + \bar{n}_{dc}} . \notag
\end{align}

Note that since we have chosen to include the dark counts in the memory transformation, attention should be payed to keeping the dark count rate fixed when photon loss ($p_{con}>0$) is introduced.

\section{Mathematical details of the recurrence example}
\label{app.recexmath}

To solve the recurrence equations \eqnref{eq.recexequations}, make the variable substitution $\tilde{f}_n = f_n^{-1}$ to obtain:
\begin{equation}
\label{eq.recexfnew}
\tilde{f}_{n+1} = 2 \tilde{f}_n - 1 .
\end{equation}
This equation is easily solved, subject to the initial condition $\tilde{f}_0 = f_0^{-1} = 2$, and we find:
\begin{equation}
\label{eq.recexfsolution}
\tilde{f}_n = 2^n + 1.
\end{equation}
Inserting the solution for $f_n$ into the $g_n$-recurrence and making the substitution $\tilde{g}_n = 2(2^n + 1) g_n$ one finds the recurrence equation
\begin{equation}
\label{eq.recexgnew}
\tilde{g}_{n+1} = 2 \tilde{g}_n - 2^{3n+2} + 2^{2n+2} - 3 .
\end{equation}
Given the initial condition $\tilde{g}_0 = 2(2^n+1)g_0 = 0$, we get the solution
\begin{equation}
\label{eq.recexgsolution}
\tilde{g}_n = - \frac{1}{3}(2^n - 1)(2^{2n+1} - 2^{n+2} - 9) .
\end{equation}
From \eqnref{eq.recexfsolution}, \eqnref{eq.recexgsolution} and the definitions of $\tilde{f}_n,\tilde{g}_n$ we have the solutions \eqnref{eq.recexsolution}.

Having found $f_n,g_n$ we may obtain the conditional Bell parameter of the state \eqnref{eq.anzats} as a function of entanglement distance. The conditional Bell parameter of the state is
\begin{align}
\label{eq.recexconfidel1}
S = 2 \sqrt{2} (f_n&-c_1^2 g_n)^2/[f_n^2 - (2 f_n g_n-(2f_n-1)^2) c_1^2  \notag \\
                   &-((2f_n-1+g_n)^2-2 g_n^2) c_1^4] .
\end{align}
Now, if $c_1$ is small such that $f_n^2$ dominates the denominator, we can plug in the solutions for $f_n,g_n$ and expand to lowest order in $c_1$. This gives the expression \eqnref{eq.recexfidel}.

The derivation of the reduction in Bell parameter in the presence of fibre loss, and due to other error sources (other memory imperfections, dark counts and finite initial squeezing)  proceed along the same lines as the derivation presented here and in \secref{sec.methods}. However, the recurrence equations tend to be considerably more complicated when losses are included, and in some cases we have not been able to obtain a closed form analytical solution. In those cases we have obtained an exact solution of the recurrence numerically (by substituting the equation into itself) and from this solution we have deduced the behaviour at large $L/L_0$. Subsequently we have verified, by comparing with numerical simulations, that the analytical expressions thus obtained are also valid for $L$ close to $L_0$.


\section{The DLCZ-repeater rate with arbitrary losses}
\label{app.dlczrate}

\begin{figure}
\centering
\includegraphics[width=.4\textwidth]{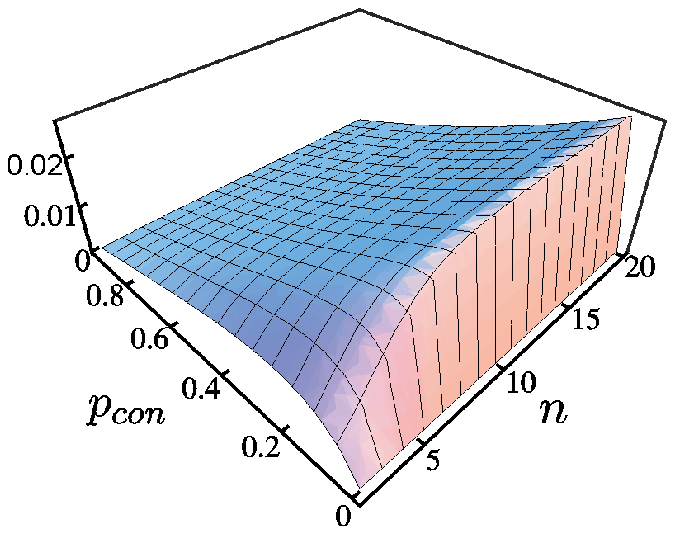}
\caption{Plot of the relative error $(\text{LHS}-\text{RHS})/\text{LHS}$ where LHS and RHS refer to \eqnref{eq.prodbestest}.}\label{fig.relerr}
\end{figure}

We consider entanglement generation and connection as shown in \figref{fig.entgenandcon}, and concentrate on the case where the photon production rate in generation is very low and where the memories are passive, such that no multi-photon errors are present. For such a system, entanglement generation produces the state $\rho_0 = \ket{\Psi^{+}}\bra{\Psi^{+}}$. Since photon loss is the only error, we expect $\rho_n$ to take the form
\begin{equation}
\label{eq.translossstate}
\rho_n = \eta_n \ket{\Psi^{+}}\bra{\Psi^{+}} + (1-\eta_n) \vacr\vacl ,
\end{equation}
where $\eta_n$ is a number and $\eta_0 = 1$. If this form of $\rho_n$ is conserved under entanglement connection, \figref{fig.entgenandcon} (b), then it follows by induction that it is correct. Assuming that the detectors resolve single photons, it is not difficult to see that this is indeed the case, and that
\begin{equation}
\label{eq.vaccompreceqn}
\eta_{n+1} = \frac{\eta_n}{2 - \eta_n (1-p_{con})} ,
\end{equation}
which has the solution
\begin{equation}
\label{eq.vaccompsolution}
\eta_n = \frac{1}{1-p_{con} + 2^n p_{con}} .
\end{equation}
Now, from \eqnref{eq.translossstate} the success probability for connection is
\begin{align}
\label{eq.translossconprob}
q_{n+1} & = \frac{1}{2} (1-p_{con}) \eta_n (2 - \eta_n(1-p_{con})) \notag \\
        & = \frac{1}{2} (1-p_{con}) \eta_n^2/\eta_{n+1} .
\end{align}
Since $r \ll 1$, the success probability for generation is $q_0 = 2 r^2(1-p_{gen})$. The probability for successful postselection is $q_{ps}=\eta_n^2/2$ and from \eqnref{eq.rate} the rate is then
\begin{equation}
\label{eq.translossrateopen}
R = \frac{2}{3^{n+1}} r^2 (1-p_{gen}) (1-p_{con})^n \eta_0^2 \prod_{i=1}^{n} \eta_i .
\end{equation}
We now put this expression on a closed form. We start by turning the product into a sum by taking the logarithm
\begin{equation}
\label{eq.logprod}
\ln{\prod_{i=1}^{n} \eta_i } = - \sum_{i=1}^{n}{\ln{(1 - p_{con} + 2^i p_{con})}} .
\end{equation}
The sum can be estimated by taking the integral
\begin{equation}
\label{eq.sumintegral}
\int_{1}^{n+1} \ln{(1 - p_{con} + p_{con}\gamma 2^x)} dx ,
\end{equation}
where we have introduced a constant $\gamma$. By adjusting $\gamma$ we make sure that the integral agrees with the sum above in the limits where the sum can be easily evaluated. The integral gives
\begin{equation}
\label{eq.sumintegralvalue}
n \ln{(1-p_{con})} + \frac{\text{Li}_2\left(\frac{2 \gamma p_{con}}{p_{con}-1}\right) - \text{Li}_2\left(\frac{2^{n+1} \gamma p_{con}}{p_{con}-1}\right)}{\ln{2}} ,
\end{equation}
where $\text{Li}_2$ is the dilogarithm \cite{morris}. Since $\text{Li}_2(0)$ is 0, the integral equals the sum in $\eqnref{eq.logprod}$ for $p_{con}\rightarrow 0$. The sum is also easily evaluated in the limit $p_{con}\rightarrow 1$. In that case it evaluates to
\begin{equation}
\label{eq.sumunphyslimit}
\sum_{i=1}^{n} \ln 2^i = \frac{\ln{2}}{2} n(n+1).
\end{equation}
Using that $\text{Li}_2(x)$ tends to $-\pi^2/6-\ln^2(-x)/2$ for large negative values of $x$ \cite{morris}, the limit of the integral \eqnref{eq.sumintegral} is
\begin{equation}
\label{eq.intunphyslimit}
\frac{\ln 2}{2} n(n + \frac{2 \ln 2\gamma}{\ln 2}) .
\end{equation}
Hence for the limit of the integral to equal that of the sum, we require $\gamma = 1/\sqrt{2}$. Inserting $\gamma$ in \eqnref{eq.sumintegralvalue} and taking the exponential, our best estimate for the product occurring in \eqnref{eq.translossrateopen} is
\begin{equation}
\label{eq.prodbestest}
\prod_{i=1}^{n} \eta_i \approx \frac{\exp \frac{1}{\ln 2} \left( \text{Li}_2 \frac{2^{n+1/2} p_{con}}{p_{con}-1} - \text{Li}_2 \frac{2^{1/2} p_{con}}{p_{con}-1} \right) }{(1-p_{con})^{n}} ,
\end{equation}
Using this together with $L/L_0=2^n$ we obtain \eqnref{eq.translossrateclosed}. It can be verified numerically that \eqnref{eq.prodbestest} is in fact a very good approximation in our range of interest. \figref{fig.relerr} shows a plot of the relative error as a function of $n$ and $p_{con}$. For $n\leq 45$ the relative error never exceeds 3\% for any value of $p_{con}$.

For non-counting detectors, a similar derivation can be carried out with the recursion $\eta_n$ modified slightly since events where two photons reach the same detector are now accepted as successful connections.

\bibliography{repeaterperformance}

\begin{thebibliography}{21}
\expandafter\ifx\csname natexlab\endcsname\relax\def\natexlab#1{#1}\fi
\expandafter\ifx\csname bibnamefont\endcsname\relax
  \def\bibnamefont#1{#1}\fi
\expandafter\ifx\csname bibfnamefont\endcsname\relax
  \def\bibfnamefont#1{#1}\fi
\expandafter\ifx\csname citenamefont\endcsname\relax
  \def\citenamefont#1{#1}\fi
\expandafter\ifx\csname url\endcsname\relax
  \def\url#1{\texttt{#1}}\fi
\expandafter\ifx\csname urlprefix\endcsname\relax\def\urlprefix{URL }\fi
\providecommand{\bibinfo}[2]{#2}
\providecommand{\eprint}[2][]{\url{#2}}

\bibitem[{\citenamefont{Briegel et~al.}(1998)\citenamefont{Briegel, D\"ur,
  Cirac, and Zoller}}]{briegel}
\bibinfo{author}{\bibfnamefont{H.-J.} \bibnamefont{Briegel}},
  \bibinfo{author}{\bibfnamefont{W.}~\bibnamefont{D\"ur}},
  \bibinfo{author}{\bibfnamefont{J.~I.} \bibnamefont{Cirac}}, \bibnamefont{and}
  \bibinfo{author}{\bibfnamefont{P.}~\bibnamefont{Zoller}},
  \bibinfo{journal}{Phys. Rev. Lett.} \textbf{\bibinfo{volume}{81}},
  \bibinfo{pages}{5932} (\bibinfo{year}{1998}).

\bibitem[{\citenamefont{Duan et~al.}(2001)\citenamefont{Duan, Lukin, Cirac, and
  Zoller}}]{dlcz}
\bibinfo{author}{\bibfnamefont{L.-M.} \bibnamefont{Duan}},
  \bibinfo{author}{\bibfnamefont{M.~D.} \bibnamefont{Lukin}},
  \bibinfo{author}{\bibfnamefont{J.~I.} \bibnamefont{Cirac}}, \bibnamefont{and}
  \bibinfo{author}{\bibfnamefont{P.}~\bibnamefont{Zoller}},
  \bibinfo{journal}{Nature} \textbf{\bibinfo{volume}{414}},
  \bibinfo{pages}{413} (\bibinfo{year}{2001}).

\bibitem[{\citenamefont{Chen et~al.}(2007)\citenamefont{Chen, Zhao, Chen,
  Schmiedmayer, and Pan}}]{chen}
\bibinfo{author}{\bibfnamefont{Z.-B.} \bibnamefont{Chen}},
  \bibinfo{author}{\bibfnamefont{B.}~\bibnamefont{Zhao}},
  \bibinfo{author}{\bibfnamefont{Y.-A.} \bibnamefont{Chen}},
  \bibinfo{author}{\bibfnamefont{J.}~\bibnamefont{Schmiedmayer}},
  \bibnamefont{and} \bibinfo{author}{\bibfnamefont{J.-W.} \bibnamefont{Pan}},
  \bibinfo{journal}{Phys. Rev. A} \textbf{\bibinfo{volume}{76}},
  \bibinfo{pages}{022329} (\bibinfo{year}{2007}).

\bibitem[{\citenamefont{Jiang et~al.}(2007)\citenamefont{Jiang, Taylor, and
  Lukin}}]{jiang}
\bibinfo{author}{\bibfnamefont{L.}~\bibnamefont{Jiang}},
  \bibinfo{author}{\bibfnamefont{J.~M.} \bibnamefont{Taylor}},
  \bibnamefont{and} \bibinfo{author}{\bibfnamefont{M.~D.} \bibnamefont{Lukin}},
  \bibinfo{journal}{Phys. Rev. A} \textbf{\bibinfo{volume}{76}},
  \bibinfo{pages}{012301} (\bibinfo{year}{2007}).

\bibitem[{\citenamefont{Simon et~al.}(2007)\citenamefont{Simon, de~Riedmatten,
  Afzelius, Sangouard, Zbinden, and Gisin}}]{simon}
\bibinfo{author}{\bibfnamefont{C.}~\bibnamefont{Simon}},
  \bibinfo{author}{\bibfnamefont{H.}~\bibnamefont{de~Riedmatten}},
  \bibinfo{author}{\bibfnamefont{M.}~\bibnamefont{Afzelius}},
  \bibinfo{author}{\bibfnamefont{N.}~\bibnamefont{Sangouard}},
  \bibinfo{author}{\bibfnamefont{H.}~\bibnamefont{Zbinden}}, \bibnamefont{and}
  \bibinfo{author}{\bibfnamefont{N.}~\bibnamefont{Gisin}},
  \bibinfo{journal}{Phys. Rev. Lett.} \textbf{\bibinfo{volume}{98}},
  \bibinfo{pages}{190503} (\bibinfo{year}{2007}).

\bibitem[{\citenamefont{Sangouard et~al.}(2007)\citenamefont{Sangouard, Simon,
  Minar, Zbinden, de~Riedmatten, and Gisin}}]{sangouard}
\bibinfo{author}{\bibfnamefont{N.}~\bibnamefont{Sangouard}},
  \bibinfo{author}{\bibfnamefont{C.}~\bibnamefont{Simon}},
  \bibinfo{author}{\bibfnamefont{J.}~\bibnamefont{Minar}},
  \bibinfo{author}{\bibfnamefont{H.}~\bibnamefont{Zbinden}},
  \bibinfo{author}{\bibfnamefont{H.}~\bibnamefont{de~Riedmatten}},
  \bibnamefont{and} \bibinfo{author}{\bibfnamefont{N.}~\bibnamefont{Gisin}}
  (\bibinfo{year}{2007}), \eprint{arXiv:0706.1924v1}.

\bibitem[{\citenamefont{Sangouard et~al.}(2008)\citenamefont{Sangouard, Simon,
  Zhao, Chen, de~Riedmatten, Pan, and Gisin}}]{sangouard2}
\bibinfo{author}{\bibfnamefont{N.}~\bibnamefont{Sangouard}},
  \bibinfo{author}{\bibfnamefont{C.}~\bibnamefont{Simon}},
  \bibinfo{author}{\bibfnamefont{B.}~\bibnamefont{Zhao}},
  \bibinfo{author}{\bibfnamefont{Y.-A.} \bibnamefont{Chen}},
  \bibinfo{author}{\bibfnamefont{H.}~\bibnamefont{de~Riedmatten}},
  \bibinfo{author}{\bibfnamefont{J.-W.} \bibnamefont{Pan}}, \bibnamefont{and}
  \bibinfo{author}{\bibfnamefont{N.}~\bibnamefont{Gisin}}
  (\bibinfo{year}{2008}), \eprint{arXiv:0802.1475v1}.

\bibitem[{\citenamefont{Chanelière et~al.}(2005)\citenamefont{Chanelière,
  Matsukevich, Jenkins, Lan, Kennedy, and Kuzmich}}]{chaneliere}
\bibinfo{author}{\bibfnamefont{T.}~\bibnamefont{Chanelière}},
  \bibinfo{author}{\bibfnamefont{D.~N.} \bibnamefont{Matsukevich}},
  \bibinfo{author}{\bibfnamefont{S.~D.} \bibnamefont{Jenkins}},
  \bibinfo{author}{\bibfnamefont{S.-Y.} \bibnamefont{Lan}},
  \bibinfo{author}{\bibfnamefont{T.~A.~B.} \bibnamefont{Kennedy}},
  \bibnamefont{and} \bibinfo{author}{\bibfnamefont{A.}~\bibnamefont{Kuzmich}},
  \bibinfo{journal}{Nature} \textbf{\bibinfo{volume}{438}},
  \bibinfo{pages}{833} (\bibinfo{year}{2005}).

\bibitem[{\citenamefont{Chou et~al.}(2007)\citenamefont{Chou, Laurat, Deng,
  Choi, de~Riedmatten, Felinto, and Kimble}}]{chou}
\bibinfo{author}{\bibfnamefont{C.-W.} \bibnamefont{Chou}},
  \bibinfo{author}{\bibfnamefont{J.}~\bibnamefont{Laurat}},
  \bibinfo{author}{\bibfnamefont{H.}~\bibnamefont{Deng}},
  \bibinfo{author}{\bibfnamefont{K.~S.} \bibnamefont{Choi}},
  \bibinfo{author}{\bibfnamefont{H.}~\bibnamefont{de~Riedmatten}},
  \bibinfo{author}{\bibfnamefont{D.}~\bibnamefont{Felinto}}, \bibnamefont{and}
  \bibinfo{author}{\bibfnamefont{H.~J.} \bibnamefont{Kimble}},
  \bibinfo{journal}{Science} \textbf{\bibinfo{volume}{316}},
  \bibinfo{pages}{1316} (\bibinfo{year}{2007}).

\bibitem[{\citenamefont{Eisaman et~al.}(2005)\citenamefont{Eisaman, André,
  Massou, Fleischhauer, Zibrov, and Lukin}}]{eisaman}
\bibinfo{author}{\bibfnamefont{M.~D.} \bibnamefont{Eisaman}},
  \bibinfo{author}{\bibfnamefont{A.}~\bibnamefont{André}},
  \bibinfo{author}{\bibfnamefont{F.}~\bibnamefont{Massou}},
  \bibinfo{author}{\bibfnamefont{M.}~\bibnamefont{Fleischhauer}},
  \bibinfo{author}{\bibfnamefont{A.~S.} \bibnamefont{Zibrov}},
  \bibnamefont{and} \bibinfo{author}{\bibfnamefont{M.~D.} \bibnamefont{Lukin}},
  \bibinfo{journal}{Nature} \textbf{\bibinfo{volume}{438}},
  \bibinfo{pages}{837} (\bibinfo{year}{2005}).

\bibitem[{\citenamefont{Chen et~al.}(2008)\citenamefont{Chen, Chen, Yuan, Zhao,
  Chuu, Schmiedmayer, and Pan}}]{chen2}
\bibinfo{author}{\bibfnamefont{Y.-A.} \bibnamefont{Chen}},
  \bibinfo{author}{\bibfnamefont{S.}~\bibnamefont{Chen}},
  \bibinfo{author}{\bibfnamefont{Z.-S.} \bibnamefont{Yuan}},
  \bibinfo{author}{\bibfnamefont{B.}~\bibnamefont{Zhao}},
  \bibinfo{author}{\bibfnamefont{C.-S.} \bibnamefont{Chuu}},
  \bibinfo{author}{\bibfnamefont{J.}~\bibnamefont{Schmiedmayer}},
  \bibnamefont{and} \bibinfo{author}{\bibfnamefont{J.-W.} \bibnamefont{Pan}},
  \bibinfo{journal}{Nature Physics} \textbf{\bibinfo{volume}{4}},
  \bibinfo{pages}{103} (\bibinfo{year}{2008}).

\bibitem[{\citenamefont{Muschik et~al.}(2006)\citenamefont{Muschik, Hammerer,
  Polzik, and Cirac}}]{muschik}
\bibinfo{author}{\bibfnamefont{C.~A.} \bibnamefont{Muschik}},
  \bibinfo{author}{\bibfnamefont{K.}~\bibnamefont{Hammerer}},
  \bibinfo{author}{\bibfnamefont{E.~S.} \bibnamefont{Polzik}},
  \bibnamefont{and} \bibinfo{author}{\bibfnamefont{J.~I.} \bibnamefont{Cirac}},
  \bibinfo{journal}{Phys. Rev. A} \textbf{\bibinfo{volume}{73}},
  \bibinfo{pages}{062329} (\bibinfo{year}{2006}).

\bibitem[{\citenamefont{Julsgaard et~al.}(2004)\citenamefont{Julsgaard,
  Sherson, Cirac, Fiuráek, and Polzik}}]{julsgaard}
\bibinfo{author}{\bibfnamefont{B.}~\bibnamefont{Julsgaard}},
  \bibinfo{author}{\bibfnamefont{J.}~\bibnamefont{Sherson}},
  \bibinfo{author}{\bibfnamefont{J.~I.} \bibnamefont{Cirac}},
  \bibinfo{author}{\bibfnamefont{J.}~\bibnamefont{Fiuráek}}, \bibnamefont{and}
  \bibinfo{author}{\bibfnamefont{E.~S.} \bibnamefont{Polzik}},
  \bibinfo{journal}{Nature} \textbf{\bibinfo{volume}{432}},
  \bibinfo{pages}{482} (\bibinfo{year}{2004}).

\bibitem[{\citenamefont{Collins et~al.}(2007)\citenamefont{Collins, Jenkins,
  Kuzmich, and Kennedy}}]{collins}
\bibinfo{author}{\bibfnamefont{O.~A.} \bibnamefont{Collins}},
  \bibinfo{author}{\bibfnamefont{S.~D.} \bibnamefont{Jenkins}},
  \bibinfo{author}{\bibfnamefont{A.}~\bibnamefont{Kuzmich}}, \bibnamefont{and}
  \bibinfo{author}{\bibfnamefont{T.~A.~B.} \bibnamefont{Kennedy}},
  \bibinfo{journal}{Phys. Rev. Lett.} \textbf{\bibinfo{volume}{98}},
  \bibinfo{pages}{060502} (\bibinfo{year}{2007}).

\bibitem[{\citenamefont{Aspect et~al.}(1982)\citenamefont{Aspect, Grangier, and
  Roger}}]{aspect}
\bibinfo{author}{\bibfnamefont{A.}~\bibnamefont{Aspect}},
  \bibinfo{author}{\bibfnamefont{P.}~\bibnamefont{Grangier}}, \bibnamefont{and}
  \bibinfo{author}{\bibfnamefont{G.}~\bibnamefont{Roger}},
  \bibinfo{journal}{Phys. Rev. Lett.} \textbf{\bibinfo{volume}{49}},
  \bibinfo{pages}{91} (\bibinfo{year}{1982}).

\bibitem[{\citenamefont{Muschik and Hammerer}()}]{privcommuschik}
\bibinfo{author}{\bibfnamefont{C.}~\bibnamefont{Muschik}} \bibnamefont{and}
  \bibinfo{author}{\bibfnamefont{K.}~\bibnamefont{Hammerer}},
  \emph{\bibinfo{title}{Private communication}}.

\bibitem[{\citenamefont{Kuzmich et~al.}(2000)\citenamefont{Kuzmich, Mandel, and
  Bigelow}}]{kuzmich}
\bibinfo{author}{\bibfnamefont{A.}~\bibnamefont{Kuzmich}},
  \bibinfo{author}{\bibfnamefont{L.}~\bibnamefont{Mandel}}, \bibnamefont{and}
  \bibinfo{author}{\bibfnamefont{N.~P.} \bibnamefont{Bigelow}},
  \bibinfo{journal}{Phys. Rev. Lett.} \textbf{\bibinfo{volume}{85}},
  \bibinfo{pages}{1594} (\bibinfo{year}{2000}).

\bibitem[{\citenamefont{Geremia et~al.}(2004)\citenamefont{Geremia, Stockton,
  and Mabuchi}}]{geremia}
\bibinfo{author}{\bibfnamefont{J.}~\bibnamefont{Geremia}},
  \bibinfo{author}{\bibfnamefont{J.~K.} \bibnamefont{Stockton}},
  \bibnamefont{and} \bibinfo{author}{\bibfnamefont{H.}~\bibnamefont{Mabuchi}},
  \bibinfo{journal}{Science} \textbf{\bibinfo{volume}{304}},
  \bibinfo{pages}{270} (\bibinfo{year}{2004}).

\bibitem[{\citenamefont{Fernholz et~al.}(2008)\citenamefont{Fernholz, Krauter,
  Jensen, Sherson, Soerensen, and Polzik}}]{fernholz}
\bibinfo{author}{\bibfnamefont{T.}~\bibnamefont{Fernholz}},
  \bibinfo{author}{\bibfnamefont{H.}~\bibnamefont{Krauter}},
  \bibinfo{author}{\bibfnamefont{K.}~\bibnamefont{Jensen}},
  \bibinfo{author}{\bibfnamefont{J.~F.} \bibnamefont{Sherson}},
  \bibinfo{author}{\bibfnamefont{A.~S.} \bibnamefont{Soerensen}},
  \bibnamefont{and} \bibinfo{author}{\bibfnamefont{E.~S.} \bibnamefont{Polzik}}
  (\bibinfo{year}{2008}), \eprint{arXiv:0802.2876v1}.

\bibitem[{\citenamefont{C.C.Gerry and P.L.Knight}(2005)}]{gerry}
\bibinfo{author}{\bibnamefont{C.C.Gerry}} \bibnamefont{and}
  \bibinfo{author}{\bibnamefont{P.L.Knight}},
  \emph{\bibinfo{title}{Introductory Quantum Optics (p49)}}
  (\bibinfo{publisher}{Cambridge University Press}, \bibinfo{address}{UK},
  \bibinfo{year}{2005}).

\bibitem[{\citenamefont{Morris}(1979)}]{morris}
\bibinfo{author}{\bibfnamefont{R.}~\bibnamefont{Morris}},
  \bibinfo{journal}{Math. Comp.} \textbf{\bibinfo{volume}{33}},
  \bibinfo{pages}{778} (\bibinfo{year}{1979}).

\end{thebibliography}

\end{document}